\documentclass[aps,twocolumn,showpacs,preprintnumbers,nofootinbib,prd,superscriptaddress,groupedaddress,10pt]{revtex4-2}

\makeatletter
\def\l@subsubsection#1#2{}
\def\l@subsubsubsection#1#2{}
\makeatother

\usepackage{graphicx,amssymb,amsmath,amsthm,amsfonts,epsfig,epsf,fixmath}
\usepackage[usenames]{color}
\usepackage{epstopdf}
\usepackage{mathtools}
\usepackage{relsize}
\usepackage{placeins} 
\usepackage{booktabs} 
\usepackage{siunitx} 
\DeclareSIUnit \parsec {pc} 
\usepackage{aas_macros}
\usepackage{bm}
\usepackage{dcolumn}
\usepackage{latexsym}
\usepackage{rotating}
\usepackage{longtable} 

\usepackage{enumerate}
\usepackage{tensor,multirow}
\usepackage{url}
\usepackage[linktocpage]{hyperref}

\def\nn{\nonumber}

\renewcommand{\Re}{\mathrm{Re}\,} 
\renewcommand{\Im}{\mathrm{Im}\,} 

\newcommand{\dd}{\mathrm{d}}

\newcommand{\totder}[2]{\frac{\mathrm{d} #1}{\mathrm{d} #2}}
\newcommand{\modulo}[1]{\left|#1\right|}

\newcommand{\nswsh}{\!\prescript{}{-2}{S^{\hat{a}\hat{\omega}}_{\ell m}}} 
\newcommand{\cswsh}{\!\prescript{}{-2}{S^c_{\ell m}}} 
\newcommand{\rswsh}{S^{\hat{a}\hat{\omega}}_{\ell m}} 
\newcommand{\Rin}{R^{\textup{in}}_{\ell m \hat{\omega}}} 
\newcommand{\Rup}{R^{\textup{up}}_{\ell m\hat{\omega}}} 

\begin{document}
\title{Assessing the detectability of the secondary spin in extreme mass-ratio inspirals with fully-relativistic numerical waveforms}
\author{
Gabriel Andres Piovano$^1$,
Richard Brito$^1$,
Andrea Maselli$^{2,3}$,
Paolo Pani$^{1}$}

\affiliation{$^{1}$ Dipartimento di Fisica, ``Sapienza" Università di Roma \& Sezione INFN Roma1, Piazzale Aldo Moro 5, 
00185, Roma, Italy}
\affiliation{$^{2}$ Gran Sasso Science Institute (GSSI), I-67100 L’Aquila, Italy}
\affiliation{$^{3}$ INFN, Laboratori Nazionali del Gran Sasso, I-67100 Assergi, Italy}

\begin{abstract} 
Extreme mass-ratio inspirals~(EMRIs) detectable by the Laser Inteferometric Space Antenna~(LISA) are unique probes of astrophysics and fundamental physics. 
Parameter estimation for these sources is challenging, especially because the waveforms are long, complicated, known only numerically, and slow to compute in the most relevant regime, where the dynamics is relativistic.
We perform a time-consuming Fisher-matrix error analysis of the EMRI parameters using fully-relativistic numerical waveforms to leading order in an adiabatic expansion on a Kerr background, taking into account the motion of the LISA constellation, higher harmonics, and also including the leading correction from the spin of the secondary in the post-adiabatic approximation. 
We pay particular attention to the convergence of the numerical derivatives in the Fisher matrix and to the numerical stability of the covariance matrix, which for some systems requires computing the numerical waveforms with approximately $90$-digit precision.
Our analysis confirms previous results (obtained with approximated but much more computationally efficient waveforms) for the measurement errors on the binary's parameters. We also show that the inclusion of higher harmonics improves the errors on the luminosity distance and on the orbital angular momentum angles by one order and two orders of magnitude, respectively, which might be useful to identify the environments where EMRIs live. 
We particularly focus on the measurability of the spin of the secondary, confirming that it cannot be measured with sufficient accuracy. However, due to correlations, its inclusion in the waveform model can deteriorate the accuracy on the measurements of other parameters by orders of magnitude, unless a physically-motivated prior on the secondary spin is imposed.
\end{abstract}
\maketitle

\section{Introduction}

Gravitational-wave~(GW) observations with the future space-based Laser
Interferometer Space Antenna~(LISA) will allow us to obtain unprecedented information about new GW sources~\cite{Audley:2017drz}. Among the most promising sources that LISA is expected to observe are extreme mass-ratio inspirals
(EMRIs)~\cite{Babak:2017tow}: compact binary systems where a small compact object (henceforth dubbed \emph{secondary}) with mass $\mu \sim 1$~--~$100 M_{\odot}$ orbits a supermassive black hole~(BH) (henceforth \emph{primary}) with mass $M \sim 10^5$~--~$10^7 M_{\odot}$. Due to the small mass ratio $q\equiv \mu/M \ll 1$, these systems can last years in the LISA frequency band, performing up to $O(1/q)$ orbital cycles before the secondary object plunges. Combined with the richness of their gravitational waveform, EMRI signals will allow us to measure some of the parameters of these sources with extreme precision~\cite{Babak:2017tow}, and perform exquisite tests of gravity and of the nature of compact objects~\cite{Gair:2012nm,Barausse:2020rsu}.

Due to their small mass ratio, the dynamics and GW emission of an EMRI can be accurately computed using tools from BH perturbation theory (see e.g.~\cite{Pound:2015tma,Barack:2018yvs,Pound:2021qin} for recent reviews). In this approach, the dynamics is solved perturbatively in the mass ratio $q\ll1$ and the spacetime of the binary can be treated as being given by the supermassive BH metric plus small perturbations due to the presence of the small companion object. In addition, for very small mass ratios, the radiation-reaction timescale is much longer than the typical orbital period so that the secondary's orbital motion around the primary can be evolved in a quasi-adiabatic fashion~\cite{Hinderer:2008dm}. The effect of the secondary spin in the GW phase enters at first order in a post-adiabatic expansion, being thus suppressed by the small mass ratio~\cite{Hartl:2002ig}, but still entering at the same order in $q$ as the leading order post-adiabatic self-force corrections~\cite{Burko:2003rv,Burko:2015sqa,Warburton:2017sxk,Akcay:2019bvk}. This fact makes it important to fully understand the impact of the secondary spin when attempting to compute accurate waveforms. Indeed, accurate parameter estimation with EMRIs will require gravitational waveforms valid up to at least first post-adiabatic order~\cite{Hinderer:2008dm}.

The impact of the secondary spin on the dynamics and GW emission in EMRIs has been studied in several works (see e.g.~\cite{Mino:1995fm,Tanaka:1996ht,Saijo:1998mn,Burko:2003rv,Yunes:2010zj,Dolan:2013roa,Burko:2015sqa,Harms:2015ixa,Harms:2016ctx,Lukes-Gerakopoulos:2017vkj,Nagar:2019wrt,Chen:2019hac}). Most recently, Ref.~\cite{Warburton:2017sxk} computed relativistic waveforms for a spinning compact object in generic inspirals around a massive nonrotating BH, including all first-order in $q$ self-force effects, whereas Refs.~\cite{Piovano:2020ooe,Piovano:2020zin} computed GW fluxes for a spinning secondary orbiting a spinning massive BH for bound circular, equatorial orbits. This was extended to eccentric, equatorial orbits in Refs.~\cite{Skoupy:2021asz,Skoupy:2021iwb}.

In practice, however, due to the complexity and the slow generation of EMRI waveforms computed using BH perturbation theory, almost all parameter-estimation studies done so far made use of approximated --~but fast to generate~-- waveforms~\cite{Barack:2003fp,Babak:2017tow,Huerta:2011kt,Huerta:2011zi,Speri:2021psr} (commonly known as ``kludge'' waveforms~\cite{Barack:2003fp,Babak:2006uv,Chua:2017ujo}). In fact, techniques to generate fast and fully relativistic EMRI waveforms have only recently started to be developed~\cite{Chua:2020stf,Hughes:2021exa,Katz:2021yft}, but so far fully Bayesian studies with these waveforms have only been done for a nonspinning secondary in eccentric orbits around a Schwarzschild massive BH~\cite{Katz:2021yft}.

Previous work~\cite{Huerta:2011kt,Huerta:2011zi} computed Fisher-matrix errors using a numerical kludge waveform including corrections due to the spin of the secondary. Their results suggest that LISA will be unable to constrain the magnitude of the secondary spin for systems with mass ratios $q \lesssim 10^{-4}$. Since the secondary spin introduces a nonnegligible dephasing~\cite{Piovano:2020ooe,Piovano:2020zin}, its unmeasurability can be probably related to correlations among the waveform parameters. One of the main purposes of this paper is study whether these conclusions hold when considering more accurate (albeit much slower to generate) waveforms. Indeed, it is known that using kludge waveforms may lead to large systematic errors when performing parameter estimation~\cite{Katz:2021yft}.

Using the methods recently developed in Refs.~\cite{Piovano:2020ooe,Piovano:2020zin}, and focusing on circular and equatorial orbits, we extend previous work by performing Fisher-error analyses using fully-relativistic waveforms computed within an adiabatic approximation but taking into account the leading-order post-adiabatic correction due to the secondary spin.
To the best of our knowledge, even neglecting the secondary spin, ours is among the first studies presenting a Fisher-matrix analysis on the EMRI parameters using fully-relativistic, Teukolsky-based waveforms on a Kerr background. The only exception is Ref.~\cite{Burke:2020vvk} where a Fisher-matrix analysis using Teukolsky-based waveforms for a nonspinning secondary and without including LISA's antenna pattern functions in the analysis, was presented. Our work should be seen as a benchmark for fully Bayesian parameter estimation studies and for other analyses using approximated (but significantly more efficient) waveforms. 

The rest of this paper is organized as follows. 
In Sec.~\ref{sec:setup} we summarize our setup and the procedure to obtain fully-relativistic, gravitational waveforms to leading order in an adiabatic expansion, also including the leading correction from the spin of the secondary in the post-adiabatic approximation.
In Sec.~\ref{sec:fisher} we explain the procedure to perform an accurate Fisher-matrix analysis for this system.
In Sec.~\ref{sec:results} we present and discuss our results (the busy reader mainly interested in the numerical results of our paper may jump directly to this section). We conclude in Sec.~\ref{sec:conclusion} with possible extensions.
Finally, we present some technical details in the appendices. Appendix~\ref{sec:app HST eq} is devoted to the resolution of Teukolsky equation in hyperboloidal-slicing coordinates; in Appendix~\ref{sec:appe_lin} we give some details on the procedure to linearize the field equations to linear order in the secondary spin; whereas Appendix~\ref{app:Fisher} provides some details on how we assess the accuracy and convergence of the Fisher-matrix error analysis.
We use $G=c=1$ units throughout and the notation follows that of~\cite{Piovano:2020zin}.

\section{Setup}\label{sec:setup}

\subsection{Orbital dynamics for a spinning secondary}

If the typical size of a body is much smaller than the curvature of 
the background spacetime, the object can be approximately treated as 
a point particle equipped with an infinite tower of multipole moments. 
The latter can be determined through a suitable expansion of the 
body's stress-energy tensor $T^{\mu\nu}$ (see~\cite{Tanaka:1996ht,Dixon:1964NCim,Dixon:1978} for a detailed 
discussion). The mass $\mu$ and the intrinsic spin $S$ of the object 
are the first two moments of this series and read
\begin{align}
\mu^2 =- p^\sigma p_\sigma \, , \qquad  S = \frac{1}{2}S^{\mu\nu}S_{\mu\nu} \, ,
\end{align}
where $p^\mu$ is the object's four-momentum and 
$S^{\mu\nu}$ is the skew-symmetric spin tensor.
The motion of a spinning particle is then determined by the 
Mathisson-Papapetrou-Dixon equations :
\begin{align}
\totder{X^\mu}{\lambda}& = v^\mu \ ,\\
\nabla_{\vec{v}}p^\mu &=-\frac{1}{2}{R^{\mu}}_{\nu\alpha\beta}v^\nu S^{\alpha \beta}\ , \label{eq:2MPD}\\ 
\nabla_{\vec{v}}S^{\mu\nu} &= 2p^{[\mu}v^{\nu]} \ , \label{eq:3MPD}\\
\mu&= -p_\mu v^\mu \ ,  \label{eq:4MPD}
\end{align}
where $ \nabla_{\vec{v}} \equiv v^\mu \nabla_\mu$, $v^\mu$ is the tangent 
vector to the representative worldline $X^\mu(\lambda)$,  with $\lambda$ 
an affine parameter.  The former provide a closed set of equations once 
a spin-supplementary condition has been fixed. We choose the 
Tulczyjew-Dixon condition:
\begin{equation}
S^{\mu\nu}p_\nu=0 \ ,
\end{equation}
which guarantees that the mass $\mu$ and spin $S$ are constants of motion~\cite{Semerak:1999qc}. We introduce the dimensionless 
spin parameter $\sigma$:
\begin{equation}
 \sigma = \frac{S}{\mu M} = \chi q \ , \label{def:sigma}
\end{equation}
where $\chi= S/\mu^2$ is the reduced spin of the secondary, and 
$q=\mu/M\ll 1$ is the binary mass ratio, with $M$ and $\mu$ being 
the mass of the primary and secondary, respectively. For EMRIs, 
the parameter $|\chi|\ll 1/q$, which implies $|\sigma|\ll1$. 

In the following, we consider a Kerr background spacetime, described 
in Boyer-Lindquist coordinates by the following line element:
\begin{align}
ds^2=&-dt^2+\Sigma(\Delta^{-1}dr^2+d\theta^2)+(r^2+a^2)\sin^2\theta d\phi^2\nonumber\\
&+\frac{2Mr}{\Sigma}(a\sin^2\theta-dt)^2\ ,
\end{align}
where $\Delta=r^2-2Mr+a^2$, $\Sigma=r^2+a^2\cos^2\theta$, and $a$ 
is the spin parameter such that $\vert a\vert \leq M$. Without loss of 
generality, we assume that the specific spin $a$ of the primary is 
aligned to the $z$-axis, namely $a\geq0$. We focus on circular equatorial 
orbits with the spin of the secondary aligned (anti-aligned) to $a$, i.e. 
$S>0$ ($S<0$).  Hereafter hatted quantities refer to dimensionless variables normalized by 
$M$, namely $\widehat \Omega = M \Omega$, $\hat a = a/ M$.

The Kerr spacetime admits two integrals of motion, the (normalized) energy $\tilde E = E/\mu$ and angular momentum $\tilde J_z = J_z/(\mu M)$~\cite{Ehlers:1977}.
Since for EMRIs $|\sigma|\ll1$, we expand both 
$\tilde{E},\tilde{J}_z$ in terms of the spin parameters, 
considering linear corrections only, such that that at first 
order in $\sigma$:
\begin{align}
\tilde E &= \tilde E^0 + \sigma \tilde E^1\quad \ , \quad \tilde J_z = \tilde J_z^0 + \sigma \tilde J_z^1 \, ,
\end{align}
with 
\begin{align}
\tilde E^0 &=  \frac{\pm\hat a+ (\hat r-2)\hat r^{1/2}}{\hat r^{3/4}\Delta_\pm}\,, \\
\tilde E^1 &= \frac{(\hat a \mp \sqrt{\hat r})(3 \hat a^2\mp 4 \sqrt{\hat r}+\hat r^2)}{2 \hat r^{11/4}\Delta_\pm^3}\,,\\
\tilde J_z^0 &=  \pm\frac{\hat r^2 + \hat a^2 \mp 2 \hat a \sqrt{\hat r}}{\hat r^{3/4}\Delta_\pm} \,, \\
\tilde J_z^1 &= \frac{1}{2 \hat r^{11/4} \Delta_\pm^{3}} \Big( 3\hat a^4 \pm \sqrt{\hat r}(3 \hat r -7)(\hat a^3+3 \hat a  \hat r^{2} )\nonumber\\
&+2 \hat a^2 \hat r (\hat r +2)+ \hat r^3 (\hat r -2)(2\hat r-9)\Big)\,,
\end{align}
where $\Delta_\pm=\sqrt{\pm2\hat a+(\hat r-3)\sqrt{\hat r}}$, and 
the upper (lower) sign corresponds to prograde (retrograde) orbits~\cite{Jefremov:2015gza}. The orbital frequency $\widehat \Omega$ is 
given by
\begin{equation}
\widehat\Omega(\hat r) = \widehat \Omega^0(\hat r) + \sigma \widehat \Omega^1(\hat r)
\end{equation}
where $\widehat{\Omega}^0(\hat r)=1/(\hat a\pm\hat r^{3/2})$ is the 
Keplerian frequency for a nonspinning particle, 
and
\begin{equation}
 \widehat\Omega^1(\hat r)= -\frac{3}{2}\frac{\sqrt{\hat r}\mp \hat a}{\sqrt{\hat r}(\hat r^{3/2}\pm a)^2}\,.
\end{equation}
The orbital dynamics is completely determined by $\tilde E, \tilde J_z $  and $\widehat\Omega$ once the orbital radius $\hat r$ and the parameters $\hat a$ and $\sigma$ are specified. 

\subsection{Radiation-reaction effects and orbital evolution}

At the adiabatic level, the rate of change of the constants of motion 
$\tilde E$ and $\tilde J_z$ is related to the fluxes carried away by 
gravitational radiation. These balance laws hold at first order in 
$\sigma$ for a spinning particle, as shown in Ref.~\cite{Akcay:2019bvk}. 
A caveat remains since --~at variance with the $\chi=0$ case~\cite{Kennefick:1998ab}~-- there is no rigorous proof yet that 
circular orbit remains circular even for a spinning secondary in the 
adiabatic approximation, i.e. that
\begin{equation}
\totder{\tilde E}{\hat t} = \widehat{\Omega} \totder{\tilde J_z}{\hat t } \, ,\label{eq:balance_eq1}
\end{equation}
holds for a spinning secondary. In principle, given a circular geodesic, 
small perturbations induced by the spin can induce 
eccentricity~\cite{Bini:2013uwa} or push the orbit off the 
equatorial plane for not aligned spins~\cite{Bini:2014soa,Mashhoon:2006fj}. 
Nevertheless, we shall assume that a circular orbit remains 
circular under radiation-reaction effects even when the secondary 
is spinning (with the spin vector (anti)aligned to the primary spin).
In this framework the energy fluxes can be expanded as well 
in $\sigma$:
\begin{equation}
\mathcal{F}(\hat r,\widehat\Omega) = \mathcal{F} ^0(\hat r,\widehat\Omega^0)  + \sigma \mathcal{F}^1(\hat r,\widehat\Omega^0,\widehat\Omega^1) \, ,
\end{equation}
at fixed spins $\hat a$ and orbital radius $\hat r$, with
\begin{align}
\mathcal{F} 
&= \frac{1}{q}\Bigg[ \bigg( \frac{\dd \tilde E}{ \dd \hat{t}} \bigg)^{\!\!H}_{\!\!\text{GW}} + \bigg( \frac{\dd 
\tilde E}{ \dd \hat{t}} \bigg)^{\!\!\infty}_{\!\!\text{GW}} \Bigg] \ ,
\end{align}
where $\left(\frac{\dd \tilde E}{ \dd \hat{t}} \right)^{\!\!H,\infty}_{\!\!\text{GW}}$ are the energy flux across the horizon and at infinity, respectively.
Let us define
\begin{equation}
\mathcal{G}(\hat r,\widehat\Omega) \coloneqq \bigg( \totder{\tilde E}{\hat{r}}\bigg)^{\!-1} \mathcal{F}(\hat r,\widehat\Omega)\ ,
\end{equation}
then, at first order in $\sigma$
\begin{align}
\mathcal{G}(\hat r,\widehat\Omega) &= \mathcal{G} ^0(\hat r,\widehat\Omega^0)  + \sigma \mathcal{G}^1(\hat r,\widehat\Omega^0,\widehat\Omega^1)\, ,\\
\mathcal{G} ^0 &=\bigg( \totder{\tilde E^0}{\hat{r}}\bigg)^{\!-1} \mathcal{F}^0 \, ,\\
\mathcal{G} ^1 &=\bigg( \totder{\tilde E^0}{\hat r}\bigg)^{\!-1} \mathcal{F}^1 - \bigg( \totder{\tilde E^0}{\hat r}\bigg)^{\!-2} \bigg(\totder{\tilde E^1}{\hat{r}}\bigg)\mathcal{F}^0\,,
\end{align}
which yield for the time evolution of the orbital radius
\begin{equation}
\totder{\hat{r}}{\hat t} = - \mathcal{G} ^0(\hat r,\widehat\Omega^0)  - \sigma \mathcal{G}^1(\hat r,\widehat\Omega^0,\widehat\Omega^1) \label{eq:radiusevol} \,.
\end{equation}
Finally, at first order in $\sigma$ the orbital phase is 
given by
\begin{equation}
\totder{\phi}{\hat t} =\widehat{\Omega}^0(\hat r) + \sigma \widehat \Omega^1(\hat r) \,. \label{eq:phaseevol}
\end{equation}
Solving Eqs.~\eqref{eq:radiusevol} and \eqref{eq:phaseevol} and linearizing them in $\sigma$ one can obtain $\hat r(\hat t)$ and $\phi(\hat t)$ to ${\cal O}(\sigma)$.

\subsection{GW fluxes in the Teukolsky formalism: linear expansion in the secondary spin}
We have computed the GW fluxes using the Teukolsky formalism.
For circular equatorial orbits, the fluxes at infinity are
\begin{align}
\bigg(\frac{\dd \tilde E}{\dd \hat t} \bigg)^{\!\infty}_{\!\text{GW}} &= 
\displaystyle\sum_{\ell=2}^{\infty}\displaystyle\sum_{m=1}^{\ell}\frac{\modulo{Z^H_{\ell m\hat{\omega}}}^2}{2\pi \hat \omega^2} = \displaystyle\sum_{\ell=2}^{\infty}\displaystyle\sum_{m=1}^{\ell}I_{\ell m}\ ,
  \label{eq:energyfluxinf}\\
\bigg(\frac{\dd \tilde J_z}{ \dd \hat{t}} \bigg)^{\!\infty}_{\!\text{GW}} &= 
\displaystyle\sum_{\ell=2}^{\infty}\displaystyle\sum_{m=1}^{\ell} 
\frac{m\!\modulo{Z^H_{\ell m \hat{\omega}}}^2}{2\pi \hat \omega^3} = \displaystyle\sum_{\ell=2}^{\infty}\displaystyle\sum_{m=1}^{\ell} 
\frac{m}{\hat \omega}I_{\ell m}\ ,
\end{align}
while at the horizon:
\begin{align}
\bigg( \frac{\dd \tilde E}{ \dd \hat{t}} \bigg)^{\!H}_{\!\text{GW}} &= 
\displaystyle\sum_{\ell=2}^{\infty}\displaystyle\sum_{m=1}^{\ell}\alpha_{\ell 
m}\frac{\modulo{Z^{\infty}_{\ell m\hat{\omega}}}^2}{2\pi\hat\omega^2} = \displaystyle\sum_{\ell=2}^{\infty}\displaystyle\sum_{m=1}^{\ell}H_{\ell m}\ , \label{eq:energyfluxhor} \\
\bigg(\frac{\dd \tilde J_z}{ \dd \hat{t}} \bigg)^{\!H}_{\!\text{GW}} &= 
\displaystyle\sum_{\ell=2}^{\infty}\displaystyle\sum_{m=1}^{\ell}\alpha_{\ell m} 
\frac{m \!\modulo{Z^{\infty}_{\ell m\hat{\omega}} }^2}{2\pi\hat\omega^3} = \displaystyle\sum_{\ell=2}^{\infty}\displaystyle\sum_{m=1}^{\ell}\frac{m}{\hat \omega}H_{\ell m}\ , \label{eq:energyfluxhor2}
\end{align}
with  $\hat \omega = m \widehat \Omega$ and the coefficient $\alpha_{\ell m} $ being 
given in~\cite{Hughes:1999bq}. 
The procedure to compute the amplitudes $I_{\ell m}$ and $H_{\ell m}$ to linear order in $\sigma$ is explained below.
By symmetry, $Z^{H,\infty}_{\ell-m-\hat{\omega}} = 
(-1)^{\ell}\bar{Z}^{H,\infty}_{\ell m\hat{\omega}}$, where the bar denotes complex 
conjugation. The complex  amplitudes 
\begin{equation}
Z^{H,\infty}_{\ell m \hat{\omega}} =Z^{H,\infty}_{\ell m \hat{\omega}}(\lambda_{\ell m\hat{\omega}} ,\nswsh,\Rin, \Rup) \ ,
\end{equation}
depend on the solutions of two decoupled ordinary differential equations, whereas $\lambda_{\ell m\hat{\omega}} $ and $\nswsh$ are respectively the eigenvalues and eigenfunctions of the angular Teukolsky equation:
\begin{eqnarray}
&&\left.\Bigg[\frac{1}{\sin\theta}\totder{}{\theta}\left(\sin\theta\totder{}{\theta} \right) - c^2\sin^2\theta - \left( \frac{m - 2 \cos\theta}{\sin\theta}\right)^{\!\!2} \right.\nn\\
&+&\left. 4 c \cos\theta- 2 
+2 mc\right.\Bigg]\cswsh = -\lambda_{\ell m\hat\omega} \cswsh  \ ,\label{eq:swsweq}
\end{eqnarray}
where $c \equiv \hat a \hat \omega$.
The following identities hold: $\lambda_{\ell m -\hat{\omega}} = \lambda_{\ell-m \hat{\omega}}$ and
\begin{equation}
 \prescript{}{-2}{S^{- c}_{\ell -m}}(\theta) = (-1)^l \!\!\, 
\prescript{}{-2}{S^c_{\ell 
m}}(\pi-\theta) \ ,
\end{equation}
while $\cswsh(\theta) e^{im\phi}$ reduces to the spin-weighted spherical harmonics for $\hat a=0$ or $\hat{\omega}=0$.  
Similarly, the functions $R^{\textup{in}}_{\ell m \hat{\omega}}$ and $R^{\textup{up}}_{\ell m \hat{\omega}}$ are linearly independent solutions of the radial Teukolsky equation:
\begin{equation}
\Delta^2 \totder{}{\hat{r}}\left(\frac{1}{\Delta}\totder{R_{\ell m\omega}}{\hat{r}}\right) - V(\hat{r}) R_{\ell m \hat{\omega}}(\hat{r}) = 0\ , \label{eq:radialTeueq}
\end{equation}
where the potential $V(\hat{r})$ reads
\begin{align}
V(\hat{r}) &= -\frac{K^2 + 4i(\hat{r}-1)K}{\Delta} + 8i \hat{\omega} \hat{r} + \lambda_{\ell m \hat{\omega}} \ 
,\\
K &= (\hat{r}^2 + \hat{a}^2)\hat{\omega} -\hat{a}m \, ,\\ 
\Delta &= \hat r^2 +\hat a^2 -2 \hat r\, ,
\end{align}
while
\begin{equation}
W_{\hat{r}} \equiv\frac{1}{\Delta}\left(\! {\Rin}\totder{\Rup}{\hat{r}} - {\Rup}{}\totder{\Rin}{\hat{r}} \!\right) \, ,
\end{equation}
is the corresponding Wronskian.
It is possible to write the amplitudes $Z^{H,\infty}_{\ell m \hat{\omega}} $ for a specific orbital radius $\hat r$ as
 \begin{align}
&Z^{H,\infty}_{\ell m \hat{\omega}} =\frac{2\pi }{W_{\hat{r}} }\left[\!A_0  - (A_1 + B_1) \totder{}{\hat{r}} + 
\right. \nonumber\\
&\left. \left.+ (A_2+B_2) \frac{\dd^2}{\dd \hat{r}^2}  - B_3 \frac{\dd^3 }{\dd \hat{r}^3} 
\right]\!R^{\textup{in},\textup{up}}_{\ell m \hat{\omega}} \right\rvert_{\theta = \pi/2 , \hat r = 
\hat r (\hat t)} \ . \label{eq:Z amp}
\end{align}
The general expressions for the coefficients $A_0, A_1, A_2$ and $B_1, B_2, B_3$, as a 
function of $\hat r$, $\lambda_{\ell m\hat{\omega}} $ and $\nswsh$, is given 
in~\cite{Piovano:2020zin}.

Following the linearized approach applied before, 
we compute spin-corrections to the fluxes~\eqref{eq:energyfluxinf} and~\eqref{eq:energyfluxhor} 
at first order in $\sigma$, keeping the orbital radius $\hat r$ fixed. 
To this aim, we first expand the solutions of the Teukolsky angular and radial 
equations, i.e.
\begin{align}
\lambda_{\ell m\hat{\omega}} &= \lambda^0_{\ell m}(c^0) + \sigma \lambda^1_{\ell m}(c^0,c^1)  \,\label{Teuklinear1} ,\\
\!\prescript{}{-2}{S^c_{\ell m}}(\theta)&= \!\prescript{}{-2}{S^0_{\ell m}}(\theta,c^0) + \sigma\!\prescript{}{-2}{S^1_{\ell m}}(\theta,c^0,c^1) \, ,\\
R^{\textup{in}}_{\ell m \hat{\omega}}(\hat r) &= R^{\textup{in},0}_{\ell m }(\hat r,\omega^0)  +\sigma R^{\textup{in},1}_{\ell m }(\hat r,\hat \omega^0,\hat \omega^1) \, ,\\
R^{\textup{up}}_{\ell m \hat{\omega}}(\hat r) &=R^{\textup{up},0}_{\ell m }(\hat r,\omega^0)  +\sigma R^{\textup{up},1}_{\ell m }(\hat r,\hat \omega^0,\hat \omega^1) \, ,\label{Teuklinear4}
\end{align}
where $\hat\omega^i = m \widehat \Omega^i$, and we expanded $c=c^0+\sigma c^1+{\cal O}(\sigma^2)$, where $c^i = \hat a \hat \omega^i$ with $i=0,1$. 
We shall now describe the procedure we adopted to compute all 
the components of Eqs.~\eqref{Teuklinear1}-\eqref{Teuklinear4} as well as 
of Eqs.~\eqref{eq:energyfluxinf}-\eqref{eq:energyfluxhor}. 

\subsubsection{Linearization in the secondary spin: Angular solutions}
If we impose regularity of the solutions at the boundaries $\theta =0$ and 
$\theta = \pi$, which are regular singular points, Eq.~\eqref{eq:swsweq} defines 
a Sturm-Liouville eigenvalue problem. Despite being a singular Sturm-Liouville 
problem (see Appendix~\ref{sec:lin angular Teu}), for real frequencies, Eq.~\eqref{eq:swsweq} 
retains much of the properties of a regular one. In particular, it can be seen 
as an eigenvalue problem for a Hermitian operator $\mathcal{H}$: 
\begin{equation}
\mathcal{H} | S \rangle = -\lambda_{\ell m\hat{\omega}}  | S \rangle \, ,
\end{equation}
where $| S\rangle \equiv \!\prescript{}{-2}{S^c_{\ell m}}(\theta)$ and $\mathcal{H}$ 
is the left-hand side of Eq.~\eqref{eq:swsweq}. If we expand $\mathcal{H}, \lambda_{\ell m\hat{\omega}}$, and $| S \rangle$ to linear order in $\sigma$, we obtain:
\begin{align}
\mathcal{H}^0 | S^0 \rangle &= -\lambda^0_{\ell m}(c^0)  | S^0 \rangle \, ,\\
\mathcal{H}^0| S^1 \rangle +\mathcal{V}^1| S^0 \rangle &= -\lambda^0_{\ell m}(c^0) | S^1 \rangle -\lambda^1_{\ell m}(c^0,c^1) | S^0 \rangle \, ,
\end{align}
where $\prescript{}{-2}{S^0_{\ell m}}(\theta,c^0) \equiv | S^0 \rangle $  and $\prescript{}{-2}{S^1_{\ell m}}(\theta,c^0,c^1) \equiv | S^1 \rangle $. The functional form of $\mathcal{V}^1$ is given in the Appendix~\ref{sec:appe_lin}, while $\mathcal{H}^0$ is simply 
given by $\mathcal{H}$ with $c \leftrightarrow c^0$.
In this fashion, we can consider $\mathcal{V}^1$ as a perturbation of an Hermitian 
operator $\mathcal{H}^0$, and  the corrections $\lambda^1_{\ell m}(c^0,c^1) $ induced by 
the spin $\sigma$ can be obtained using the same techniques of time-independent 
perturbation theory for a (nondegenerate) quantum mechanical system, i.e
\begin{equation}
\lambda^1_{\ell m}(c^0,c^1) =\langle S^0 |\mathcal{V}^1|S^0 \rangle \equiv \int_0^\pi \!\!\prescript{}{-2}{S^0_{\ell m}}\mathcal{V}^1\!\prescript{}{-2}{S^0_{\ell m}}\sin\theta\dd \theta \, .
\end{equation}

Once the corrections to the eigenvalues $\lambda^1_{\ell m}(c^0,c^1)$ are known, 
we can compute the corrections to the eigenfunctions $S^1_{\ell m}(\theta,c^0,c^1)$ by 
expanding in $\sigma$ the series coefficients of the solution obtained with 
Leaver's method (see Appendix~\ref{sec:lin angular Teu} for more details).
To compute the 0-th order eigenvalues $\lambda^0_{\ell m}(c^0)$ and eigenfunctions $\prescript{}{-2}{S^0_{\ell m}}(\theta,c^0) $ of Eq.~\eqref{eq:swsweq} we used 
Leaver's method implemented in the Black Hole Perturbation Toolkit~\cite{BHPToolkit}.

It is worth to remark that we can always find the exact solutions 
of Eq.~\eqref{eq:swsweq} for any value of $\sigma$, and then interpolate 
to extract the first order correction in the spin. However, the semi-analytic 
linearization approach described above provides a powerful and fast 
method to avoid such numerical procedure. It may happen, though, that 
in some regions of the parameter space, the input parameters require higher 
precision than expected due to large numerical cancellations in the algorithm. 
When the precision of the corrections obtained with the semi-analytic method 
dropped below a certain threshold, we used as a ``backup'' approach a simple 
interpolation from the exact solutions, i.e.
\begin{align}
\lambda^1_{\ell m} &= \frac{\lambda_{\ell m \hat \omega}(c^0 + \epsilon c^1) -\lambda_{\ell m \hat \omega}(c^0 - \epsilon c^1)}{\epsilon} \, ,\\
\prescript{}{-2}{S^1_{\ell m}} &= \frac{\prescript{}{-2}{S^{(c^0 + \epsilon c^1)}_{\ell m}} -\prescript{}{-2}{S^{(c^0 - \epsilon c^1)}_{\ell m}}}{\epsilon} \, ,
\end{align}
where the exact eigenvalues $\lambda_{\ell m \hat \omega}(c^0 + \epsilon c^1)$, $\lambda_{\ell m \hat \omega}(c^0 - \epsilon c^1)$ and eigenfunctions $\prescript{}{-2}{S^{(c^0 + \epsilon c^1)}_{\ell m}}$, $\prescript{}{-2}{S^{(c^0 -\epsilon c^1)}_{\ell m}}$ of~\eqref{eq:swsweq} were computed using the Leaver method of the Black Perturbation Toolkit with $\epsilon = 10^{-6}$. 
We have checked that the corrections obtained with the semi-analytic method and with the 
numerical interpolation agree in all the parameter space under investigation.

\subsubsection{Linearization in the secondary spin: Radial solutions}
Equation~\eqref{eq:radialTeueq} is a stiff differential equation, i.e. the solutions of physical interest are fast oscillating functions with amplitudes increasing as $\hat r^3$ at infinity. The stiffness is caused by the long range of the potential, which makes it challenging to obtain accurate solution in the domain of integration.  Two workarounds of this issue are the semi-analytic Mano-Suzuki-Takasugi method~\cite{Fujita:2009us,Fujita:2004rb} and the numerical Sasaki-Nakamura method~\cite{Sasaki:1981sx}. Here we employed a third method, which consists in considering a particular ansatz of the solutions of  Eq.~\eqref{eq:radialTeueq} based on hyperboloidal-slicing coordinates~\cite{Zenginoglu:2011jz}. Such ansatz is\footnote{The original ansatz used in~\cite{Zenginoglu:2011jz} [their Eq.~(13)] has wrong signs in some factors.}
\begin{equation}
R_{\ell m \hat \omega}(\hat r) = \hat r^{-1} \Delta^{-s} e^{\mp i \hat \omega \hat r^*}e^{i m \tilde \phi} \psi(\hat r) \ , \label{eq:ansatzHST}
\end{equation}
when the minus (plus) sign refers to $\Rin \,(\Rup) $, $s$ refers to the spin of the perturbation of the Kerr metric ($s=0,\pm 1,\pm 2$ for scalar, vector and metric perturbations, respectively), and 
\begin{align}
\tilde \phi &= \frac{\hat a}{\hat r_+ - \hat r_-}\ln\Big(\frac{\hat r - \hat r_+}{\hat r - \hat r_-}\Big)\ ,\\
\hat r^* &= \hat{r} + \frac{2 \hat{r}_+}{\hat{r}_+ - \hat{r}_-}\ln\Big(\frac{\hat{r} -\hat{r}_+}{2}\Big) - \frac{2 r_-}{r_+ -  \hat{r}_-}\ln\Big(\frac{\hat{r} - \hat{r}_-}{2}\Big)\ ,
\end{align}
with $\hat r_\pm = 1 \pm \sqrt{1 -\hat a^2}$. 
By plugging the ansatz~\eqref{eq:ansatzHST} in Eq.~\eqref{eq:radialTeueq}, we obtain an ordinary differential equation for $\psi$:
\begin{equation}
\Delta^2\frac{\dd^2\psi}{\dd \hat r^2} +\Delta\tilde F(\hat r;H)\totder{\psi}{\hat r} + \tilde U (\hat r;H)\psi = 0\ , \label{eq:radialHST}
\end{equation}
where the functions $\tilde F (\hat r;H)$ and $\tilde U (\hat r;H)$ are given in Appendix~\ref{sec:app HST eq}. Solving Eq.~\eqref{eq:radialHST} numerically is much easier than solving Eq.~\eqref{eq:radialTeueq} because the potential $\tilde U (\hat r;H)/\Delta^2$ is short ranged and the oscillating behavior at the horizon and infinity is already factored out in the ansatz~\eqref{eq:ansatzHST}. It is worth noticing that the oscillating term $ e^{\mp i \hat \omega \hat r^*}$ does not enter in the Wronskian $W_{\hat r}$. We found exact boundary conditions for Eq.~\eqref{eq:radialHST}, which allowed us to find the radial solutions $\Rin$ and $\Rup$ fast and accurately. Such boundary conditions are provided in Appendix~\ref{sec:HSC BCs}.

After expanding the ansatz~\eqref{eq:ansatzHST} as shown in Appendix~\ref{sec:lin radial Teu},
we obtained some algebraic formulas for $R^{\textup{in},1}_{\ell m}$ and $R^{\textup{up},1}_{\ell m}$ that depend on the linear corrections $\psi^{\textup{in},0},\psi^{\textup{in},1}$ and $\psi^{\textup{up},0},\psi^{\textup{up},1}$. We computed such solutions by solving a system of ordinary differential equations derived by expanding Eq.~\eqref{eq:radialHST} and the related boundary conditions to ${\cal O}(\sigma)$. See Appendix~\ref{sec:lin radial Teu} for more details.

\subsubsection{Linearization in the secondary spin: GW fluxes}
 Once the zeroth- and first-order corrections to the Teukolski variables are known, it is then possible to expand the complex amplitudes $Z^{H,\infty}_{\ell m \hat{\omega}} $ as
\begin{align}
Z^H_{\ell m \hat{\omega}}(\hat r) &= Z^{H,0}_{\ell m }(\hat r ,\omega^0)  +\sigma Z^{\textup{H},1}_{\ell m }(\hat r,\hat \omega^0,\hat \omega^1) \, ,\\
Z^\infty_{\ell m \hat{\omega}}(\hat r) &=Z^{\infty,0}_{\ell m }(\hat r,\omega^0)  +\sigma Z^{\infty,1}_{\ell m }(\hat r,\hat \omega^0,\hat \omega^1) \, ,
\end{align}

and finally obtain the correction to the fluxes at the horizon and infinity for each $\ell, m$ as follows:
\begin{align}
I_{\ell m }(\hat r) &= I^0_{\ell m }(\hat r ,\omega^0)  +\sigma I^1_{\ell m}(\hat r,\hat \omega^0,\hat \omega^1) \, , \label{eq:flux corr inf}\\
H_{\ell m }(\hat r) &= H^0_{\ell m }(\hat r ,\omega^0)  +\sigma H^1_{\ell m}(\hat r,\hat \omega^0,\hat \omega^1)\, , \label{eq:flux corr hor}
\end{align}
where $I_{\ell m }$ and $H_{\ell m }$ have been defined in Eqs.~\eqref{eq:energyfluxinf} and \eqref{eq:energyfluxhor}, respectively.
The coefficients $I^0_{\ell m }, I^1_{\ell m}$ and $H^0_{\ell m }, H^1_{\ell m}$ are given in Appendix~\ref{sec:lin source terms}.

To compute the fluxes, we constructed a nonuniform grid in the orbital radius $\hat r$  defined as follows: given $v(\hat r) \equiv (\widehat\Omega^0)^{1/3}=(\hat{r}^{3/2}+\hat a)^{-1/3}$, we considered $180$ points for $a<0.99$ and $200$ points for $a=0.99$ evenly spaced in $v$, starting from $v_{\rm start}=v(\hat r = 14)$ and ending at $v_{\rm end}=v(\hat r_{\rm{ISCO}})$, with $\hat r_{\textup{ISCO}}$ being the ISCO 
for a nonspinning test particle. The radiation reaction grid in $\hat r$ was then obtained as the solution of $\hat r_i = (1/v^3_i -\hat a)^{-2/3}$ for $i= 1, \dots 180 \,(200)$ for $\hat a<0.99$ ($\hat a=0.99$).

In the computation of the fluxes, we summed over all multipoles $\ell$ up to $\ell_{{\rm max}}=20$ ($\ell_{{\rm max}}=24$) for $a<0.99$ ($a=0.99$), summing over the index $m = 1, \dots ,\ell$ for each harmonic index $\ell$. As shown in Table~I of Ref.~\cite{Piovano:2020zin}, the fractional error in truncating the multipole sum at $\ell_{{\rm max}}$ is no larger than $\sim 10^{-5}$.

Finally, we compared the linearized fluxes with the results available in the literature. 
In the case of a Schwarzschild spacetime, our results are in perfect agreement with those of Ref.~\cite{Akcay:2019bvk} (they agree within all the digits shown in Table~I of~\cite{Akcay:2019bvk}). 
In Ref.~\cite{Piovano:2020zin}, the linear corrections to the fluxes in a Kerr spacetime were computed through a cubic interpolation of the exact fluxes in $\sigma$ (we refer to the first-order corrections computed in this way as $\mathcal F^1_{\textup{inter}}$). In order to compare with the semianalytic linear corrections $\mathcal F^1$ obtained in this work, we recomputed $\mathcal F^1_{\textup{inter}}$  as done in Ref.~\cite{Piovano:2020zin} with the following differences:
\begin{itemize}
\item we solved the radial Teukolsky equation in hyperboloidal slicing coordinates, using the same radiation-reaction grid adopted here;
\item for each $\ell$, we summed over all azimuthal indexes  $m = 1, \dots ,\ell$, as done in this work.
\end{itemize}
The fractional difference between $\mathcal F^1_{\textup{inter}}$ and $\mathcal F^1$ is, at most, $10^{-10} \%$ ($10^{-4}\%$) for $\hat a = 0.9$ ($\hat a = 0.99$) (the largest differences occurring at the ISCO).

\subsection{Waveform computation}\label{sec:waveforms}
We focus on EMRIs on circular and equatorial orbits, for which the emitted waveform in the Teukolsky formalism is given by 
\begin{align}
 h_+- i h_\times &= 2 \frac{\mu}{D} \displaystyle \sum_{\ell,m} \mathcal{A}_{\ell m \hat \omega}(t)\cswsh(\vartheta,t) e^{-i \Phi(t)}\,, \label{Teukolskywave}\\
 \Phi(t) &=  m\phi(t) +m (\varphi + \phi_0) \,,
\end{align}
where $\phi_0$ is the initial orbital phase, 
$\mathcal{A}_{\ell m \hat \omega} \equiv \hat Z^H_{\ell m \hat \omega} /\hat \omega^2$, and $\hat Z^H_{\ell m \hat \omega} = M^2 Z^H_{\ell m \hat \omega}$. $D$ is the source's luminosity distance from the detector\footnote{In this detector frame configuration, 
the component masses in Eq.~\eqref{Teukolskywave} are rescaled with 
respect to the source-frame quantities by the redshift factor $(1+z)$.}, 
and $(\vartheta,\varphi)$ identify the direction, in Boyer-Lindquist coordinates, 
of the latter in a reference frame centered at the source.
Since $\phi_0$ in Eq.~\eqref{Teukolskywave} is degenerate with the 
azimuth direction $\varphi$, from now on we will identify the 
initial phase as $\phi_0 \rightarrow \varphi + \phi_0$. 
From Eq.~\eqref{Teukolskywave} it is straightforward to identify 
the two waveform polarizations
\begin{align}
h^+_{\ell m}& = 2 \frac{\mu}{D}\cswsh (\Re \mathcal{A}_{\ell m \hat \omega}  \cos \Phi  + \Im \mathcal{A}_{\ell m \hat \omega} \sin\Phi)\, ,\\
h^\times_{\ell m} &= 2 \frac{\mu}{D}\cswsh(\Re \mathcal{A}_{\ell m \hat \omega} \sin\Phi - \Im \mathcal{A}_{\ell m \hat \omega} \cos\Phi) \, ,
\end{align}
being $\Re \mathcal{A}_{\ell m \hat \omega}$ and $\Im \mathcal{A}_{\ell m \hat \omega}$ the real and imaginary parts of $ \mathcal{A}_{\ell m \hat \omega}$.
In the presence of the secondary spin, we expand the amplitudes 
$\mathcal{A}_{\ell m \hat \omega} = \mathcal{A}_{\ell m}^0(\hat \omega^0) +\sigma \mathcal{A}_{\ell m}^1(\hat \omega^0, \hat \omega^1) + \mathcal{O}(\sigma^2)$, 
where 
\begin{align}
\mathcal{A}_{\ell m}^0& =  \frac{\hat Z^{H,0}_{\ell m \hat \omega}}{(\hat \omega^0)^2} \, ,\\
\mathcal{A}_{\ell m}^1& =  -2\frac{\hat\omega^1}{\hat\omega^0}\mathcal{A}_{\ell m}^0 + \frac{\hat Z^{H,1}_{\ell m \hat \omega}}{(\hat \omega^0)^2}  \, .
\end{align}
Therefore, we recast the two polarizations as:
\begin{align}
h^+_{\ell m}& = 2 \frac{\mu}{D} \big(\!\prescript{}{-2}{S^0_{\ell m}} + \sigma\!\prescript{}{-2}{S^1_{\ell m}} \big)A^+_{\ell m} \, ,\\
h^\times_{\ell m} &= 2 \frac{\mu}{D} \big(\!\prescript{}{-2}{S^0_{\ell m}} + \sigma\!\prescript{}{-2}{S^1_{\ell m}} \big)  A^\times_{\ell m} \, ,
\end{align}
with 
\begin{align}
A^+_{\ell m} &=   \Re\!\big( \mathcal{A}_{\ell m}^0+\sigma\mathcal{A}_{\ell m}^1 \big) \!\cos \Phi  + \Im\!\big( \mathcal{A}_{\ell m}^0+\sigma\mathcal{A}_{\ell m}^1 \big) \!\sin\Phi\, ,  \\
A^\times_{\ell m} &=   \Re\!\big( \mathcal{A}_{\ell m}^0+\sigma\mathcal{A}_{\ell m}^1 \big)\!  \sin \Phi  - \Im\!\big( \mathcal{A}_{\ell m}^0+\sigma\mathcal{A}_{\ell m}^1 \big) \!\cos\Phi \, .
\end{align}

The LISA response to the GW signal emitted by an EMRI can be written 
in terms of the $+,\times$ polarizations as 
\begin{align}
h_\alpha(t)= F^+_\alpha(\vartheta_D,&\varphi_D,\Psi)h_+(t,D,\vartheta,\varphi) + \nonumber \\
&+F^\times_\alpha(\vartheta_D,\varphi_D,\Psi)h_\times(t,D,\vartheta,\varphi) \, , \label{eq:gwsignal}
\end{align}
where $\alpha=I,II$ refers to the two independent Michelson-like detectors 
that constitute the LISA response~\cite{Gourgoulhon:2019iyu}.
The antenna pattern functions\footnote{For simplicity, we assume that $F_{+,\times}$ are constant within the frequency range 
sampled by the binary configurations considered. However, for values of $f$ 
larger than $f_\ast = 19.1\,{\rm mHz}$, LISA's antenna pattern functions also depend 
on the GW frequency~\cite{Cornish:2018dyw}.}
$F^+_\alpha$ and $F^\times_\alpha$ depend on the direction $(\vartheta_D,\varphi_D)$ of the source with respect to the detector's frame and on the polarization angle $\Psi$~\cite{Huerta:2011kt}:
\begin{align}
F_I^+=\frac{1}{2}(1+\cos^2\vartheta_D)&\cos(2\varphi_D)\cos(2\Psi) \nonumber\\
&-\cos\vartheta_D\sin(2\varphi_D)\sin(2\Psi)\ , \\
F_I^\times=\frac{1}{2}(1+\cos^2\vartheta_D)&\cos(2\varphi_D)\sin(2\Psi)\nonumber\\
&+\cos\vartheta_D\sin(2\varphi_D)\cos(2\Psi)\ ,
\end{align}
where $F_{II}^{+,\times}$ can be obtained by rotating $\varphi_D$ in the 
previous expressions by $-\pi/4$. i.e. 
$F_{II}^{+,\times}(\vartheta_D,\varphi_D,\psi)=F_I^{+,\times}(\vartheta_D,\varphi_D-\pi/4,\psi)$.

Given the LISA satellite motion, such angles are not constant but vary with time. 
However it is possible to recast $(\vartheta_D, \varphi_D,\Psi)$ in terms of fixed angles $(\vartheta_S, \varphi_S)$ and $(\vartheta_K, \varphi_K)$ which 
provide the direction of the source and of the orbital angular momentum (which for equatorial orbits coincides with the direction of the primary spin) in a  
heliocentric reference frame attached with the ecliptic~\cite{Barack:2006pq}. 
The same applies to the polar angle $\vartheta$ in the signal~\eqref{Teukolskywave}:
\begin{equation}
\cos \vartheta = \cos \vartheta_S \cos \vartheta_K + \sin \vartheta_S \sin \vartheta_K \cos(\varphi_S - \varphi_K)\ .
\end{equation}
Finally, we also include the effect of the Doppler modulation by 
introducing an offset in the phase
\begin{equation}
\Phi(t) \to \Phi(t) +  \frac{\hat \omega R}{M} \sin \vartheta_S \cos[2\pi (t/T_{{\rm LISA}})-\varphi_S] \, ,
\end{equation}
where $R= 1{\rm AU}$ and $T_{{\rm LISA}}=1\,{\rm yr}$ is LISA's orbital period~\cite{Huerta:2011kt}.

We have considered $T=1\,{\rm yr}$ observation time, ending the orbital evolution 
at the onset of the transition region as defined in~\cite{Ori:2000zn}, i.e. at $\hat r_{\textup{ISCO}}+\delta \hat r$ with $\delta \hat r = 4 q^{2/5}$. We have chosen 
$\delta \hat r$ by setting $X=1$ and $R_0=4$ in Eq.~(3.20) of~\cite{Ori:2000zn} for all the 
configurations analysed. In general, $\delta \hat r \sim \gamma q^{2/5}$ with 
$\gamma \sim O(1)$, and we checked that the Fisher matrices computed below are unaffected by the 
specific value of $\gamma$, since the signal-to-noise ratio~(SNR) accumulated around the transition region is negligible.

\section{Accurate Fisher matrix analysis for EMRI waveforms}\label{sec:fisher}

In Ref.~\cite{Piovano:2020zin} we computed the GW dephasing due 
to a nonvanishing secondary spin, showing that the effect of the secondary spin can contribute to more than $1\,{\rm rad}$ dephasing, therefore suggesting that it could provide detectable effects. However, such a simplified analysis neglects possible 
correlations between the waveform parameters that might hamper 
their measurability, especially for subleading terms. In order to 
gain a deeper insight on the detectability of the 
secondary spin in the following we shall perform a Fisher matrix analysis.

The GW signal emitted by an EMRI with a spinning 
secondary, moving on the equatorial plane with spin (anti)aligned 
to the $z$-axis, is completely specified by eleven parameters 
$\vec{x}=\{\vec{x}_\textnormal{I},\vec{x}_\textnormal{E}\}$:
(i) five intrinsic parameters $\vec{x}_\textnormal{I}=(\ln\mu, \ln M , \hat{a}, \hat{r}_0$, and $\chi$) and (ii) six extrinsic parameters
$\vec{x}_\textnormal{E}=(\phi_0,\vartheta_S, \varphi_S, \vartheta_K, \varphi_K, \ln D$), 
where we remind that: $(M,\mu)$ are the mass components with $q=\mu/M\ll1$, 
$(\hat{a},\chi)$ are the primary and secondary 
spin parameters, $(\phi_0,r_0)$ define the binary 
initial phase and orbital radius, and $D$ is the source 
luminosity distance. The four angles $(\vartheta_S,\varphi_S)$ 
and $(\vartheta_K,\varphi_K)$ correspond to the colatitude and the 
azimuth of the source sky position and of the orbital angular momentum, 
respectively~\cite{Barack:2006pq}.
Since the orbit is circular and equatorial, the orbital angular 
momentum has no precession around the primary spin, and the orbital and primary angular momenta are parallel to each other.

In the limit of large SNR, the errors on the 
source parameters inferred by a given EMRI observation can be 
determined using the Fisher information matrix:
\begin{equation}
\Gamma_{ij} =\sum_{\alpha=I,II}^{} \left(\frac{d \tilde h_\alpha}{dx^i}\middle| \frac{d \tilde h_\alpha}{d x^j}\right)_{\vec x = \vec x_0}\ ,
 \end{equation}
where $\vec x_0$ corresponds to the true set of binary parameters, 
and we have introduced the noise-weighted scalar product between two 
waveforms $p_\alpha$ and $q_\alpha$ in the frequency domain: 
\begin{equation}
(p_\alpha|q_\alpha) = 2 \int_{f_{\rm min}}^{f_{\rm max}}\frac{df}{S_n(f)}
[\tilde p^*_\alpha(f)\tilde q_\alpha(f)+\tilde p_\alpha(f)\tilde q^*_\alpha(f)]\ . \label{innerproduct}
\end{equation}
Here the tilded quantities correspond to the Fourier transform of the time-domain 
waveforms, and a star identifies complex conjugation. We used Simpson's integration rule to compute the scalar product.
As discussed in the previous section, the index $\alpha$ runs over the two independent 
channels of the LISA interferometer. 
In our computations we set $f_{\rm min}=10^{-4}\,{\rm Hz}$, while we choose $f_{\rm max}$ as
\begin{equation}
f_{\rm max}=\frac{\ell_{{\rm max}}}{2\pi}\frac{1}{M}\Big[\widehat{\Omega}^0(\hat r_{\rm ISCO}) + \sigma \widehat \Omega^1(\hat r_{{\rm ISCO}})\Big]\ ,
\end{equation}
where $\hat r_{\rm ISCO}$ is the ISCO for a nonspinning test particle and $\ell_{{\rm max}}$ the maximum harmonic index $\ell$ considered for a given system. 
Following the Shannon theorem, for the sampling time we used $\Delta t_s = \lfloor 1/(2 f_{\rm max} ) -1 \rfloor$
while the number of samples $n_s =T/\Delta t_s$ is adjusted to be an even number for a more efficient computation of the fast Fourier transform. As discussed before, for all systems the binary evolves for $T=1\,{\rm yr}$ before the plunge, so the frequency content of the signal is smaller than the range $[f_{\rm min},f_{\rm max}]$.

The waveform scalar product also allows to define the SNR for a given signal $h$ as 
\begin{equation}
{\rm SNR} = (h|h)^{1/2}\,,
\end{equation}
which scales linearly with the inverse of the luminosity 
distance. Furthermore, in the large-SNR limit the covariance 
matrix scales inversely with the SNR so, for a given set of 
parameters, it is straightforward to rescale the errors by 
changing the distance $D$ (and hence the SNR).

The inverse of $\Gamma_{ij}$ yields the covariance matrix, 
$\Sigma_{ij}$, whose diagonal elements correspond to the statistical uncertainties of the waveform parameters, 
\begin{equation}
\sigma^2_{x_i}=\Sigma_{ii}=(\Gamma^{-1})_{ii}\ ,\label{fishererrors}
\end{equation}
whereas the off-diagonal elements correspond to the correlation coefficients,
\begin{equation}
c_{x_i x_j}=\Sigma_{ij}/\sqrt{\Sigma_{ii}\Sigma_{jj}}\ .\label{fisherecorr}
\end{equation}

Hereafter we consider two data-analysis scenarios, depending 
on whether we also include a prior probability functions on the spin of the secondary or not. We follow the approach described 
in~\cite{Poisson:1995ef}, assuming for the prior a Gaussian 
distribution $p_0(\chi)$ with standard deviation 
$\sigma_\chi=1$. Given $\Gamma_0$ the Fisher matrix 
of the prior (which in our case has all vanishing elements except for the diagonal term corresponding to the secondary spin, with $(\Gamma_0)_{\chi\chi}=1/\sigma_\chi$), the new errors on the source parameters are obtained by modifying Eq.~\eqref{fishererrors} as
\begin{equation}
\sigma^2_{x_i}=[(\Gamma+\Gamma_0)^{-1}]_{ii}\ .
\end{equation}

In addition to the standard deviations on the eleven parameters 
defined above, we also analyze the error box on the solid angle 
spanned by the unit vector associated to $(\vartheta_{S},\varphi_{S})$ and $(\vartheta_{K},\varphi_{K})$:
\begin{equation}
 \Delta\Omega_{i}=2\pi |\sin\vartheta_{i}|\sqrt{\sigma^2_{\vartheta_{i}}\sigma^2_{\varphi_{i}}
 -\Sigma^2_{\vartheta_{i}\varphi_{i}}}\ .
\end{equation}
where $i=(S,K)$.

From a technical point of view, the fact that the EMRI waveform is known numerically implies that, to compute the Fisher matrix, one needs to evaluate numerical derivatives.
Apart from the derivative with respect to the luminosity distance $D$ (which can be obtained analytically since the waveform scales as $h\sim1/D$), we have computed the derivatives of the 
other ten parameters using the five-points stencil formula, namely: 
\begin{align}
\frac{d h}{d x}=\frac{1}{12 \epsilon}[h(x-2 \epsilon)-h(x&+2 \epsilon)+ 8 h(x+\epsilon)\nonumber\\
&-8h(x-\epsilon)]+\mathcal O (\epsilon^4)\ .
\end{align}
The numerical derivative is sensitive to the value of the shift 
$\epsilon$ chosen to compute the finite differences. We have 
explored various combinations of $\epsilon$ for each parameter, 
finding in general a range of at least two orders of magnitude 
in which the Fisher (and the covariance) matrices show convergence 
in the small-$\epsilon$ limit (see Appendix~\ref{app:Fisher} for a detailed analysis).

It is well known that the Fisher matrices used for the data-analysis of EMRIs are badly ill-conditioned~\cite{Vallisneri:2007ev}, which means that a small perturbation in the matrix (due to numerical or systematic errors) is greatly amplified after computing the inverse. As a rule of thumb, for a condition number\footnote{For a symmetric, positive-definite matrix, the condition number $\kappa$ is given by the ratio between the largest and the smallest of the matrix eigenvalues.} $\kappa=10^k$, one may lose up to $k$ digits of accuracy, which should be added to the numerical errors.
In our setup, an accurate inversion of the Fisher matrix requires at least $60$-digit precision in the waveform in most of the configurations, and in the worst case (namely $\hat a =0.9, \chi =1 , \mu =10,100 M_\odot$), up to $90$-digit precision. 
To achieve such precision in the waveform, we have computed 
the GW fluxes with $70$-digit precision ($100$-digit precision in the most demanding case), which allowed us to derive the Fisher matrices with no less than $38$-digit precision. 
In Appendix~\ref{app:Fisher} we provide a detailed analysis of the stability 
of the Fisher matrix for the problem at hand.

\section{Results and discussion}\label{sec:results}

\subsection{Settings}

We have computed the numerical integral in Eq.~\eqref{innerproduct} using the 
LISA noise sensitivity curve of Ref.~\cite{Cornish:2018dyw}, including the contribution 
of the confusion noise from the unresolved Galactic binaries assuming $T=1\,{\rm yr}$ of observation time.
In order to reduce the spectral leakage in the frequency domain due to the 
Fourier transform, we have tapered the time-domain waveforms with a Tukey window with 
window size $\beta = 0.05$. We checked that our results do not change noticeably when varying $\beta$ around this fiducial value.

For simplicity, in our analysis we fix the injected angles to the fiducial values $\vartheta_S = \pi/4,\phi_S=0, \vartheta_K=\pi/8, \phi_K =0$.   
Moreover, we consider a primary mass $M=10^6M_\odot$, and two choices of the secondary mass: $\mu=(10,100)M_\odot$. We compute the Fisher matrices for sources at fixed luminosity distance $D=1\,{\rm Gpc}$, but renormalize the results to a fixed fiducial SNR such that ${\rm SNR}=30$ and ${\rm SNR}=150$, for the two choices of $\mu$, respectively.

In order to analyze how the inclusion of higher-order ($\ell\geq2$) multipoles
in the signal~\eqref{eq:gwsignal} may affect the measurement of the source 
parameters, in the following we consider the purely quadrupolar case ($\ell=2$), and the cases in which the octupole ($\ell=3$) and the hexadecapole ($\ell=4$) are included.

Finally, we shall discuss two cases separately: first, in Sec.~\ref{sec:results_nospin} we neglect the spin of the secondary (i.e., removing $\chi$ from the waveform parameters); then, in Sec.~\ref{sec:results_spin} we perform a more comprehensive analysis by including also the secondary spin.

\begin{table*}[th]
\centering
\begin{tabular}{c|cccc|ccc}
\hline
\hline
$\ell$ & $\ln M$  & $\ln \mu$ &$\hat r_0$ & $\phi_0$ & $\ln D$ & $\Delta \Omega_S$ & $\Delta \Omega_K$\\
  \hline
2 & -4.62  &-4.19 &  -4.96 & 0.54 &-0.27 & $3.1\times 10^{-3}$ & 1.5 \\
2+3 &-4.64  &-4.22 &  -4.97 & -0.66 &-1.46 & $2.4\times 10^{-3}$  &$7.9\times 10^{-3}$\\
2+3+4 &-4.64  &-4.22 &  -4.97 & -0.67 &-1.46 & $2.4\times 10^{-3}$ & $7.3\times 10^{-3}$ \\
  \hline
   \hline
\end{tabular}
\caption{Errors on the intrinsic source parameters, on the luminosity distance, and 
on the solid angles which define the orientation and the orbital angular momentum of the binary, for various choices of the multipoles included in the waveform. Both EMRI components are nonspinning ($\hat{a}=\chi=0$), with $M = 10^6 M_\odot$ and $\mu = 10 M_\odot$. 
We neglect the spin parameters of both binary components ($\hat{a}$ and $\chi$) in the waveform.
The SNR for the three configurations ($D=1\,{\rm Gpc}$) is ${\rm SNR}=(22.2,24.8,25.2)$, but the errors are all normalized to the fiducial value ${\rm SNR}=30$.
For clarity, we present the $\log_{10}$ of the errors on $\ln M$, $\ln \mu$, $\hat r_0$, $\phi_0$, and $\ln D$. For example, an entry ``$-4$" for $\ln M$ ($\hat r_0$) means that the relative (absolute) error on $M$ ($\hat r_0$) is $10^{-4}$.
}\label{tab:Schwarzschilderrnospin}
\end{table*}

\begin{table*}[th]
\centering
\begin{tabular}{c|ccccc|ccc}
\hline
\hline
$\ell$ & $\ln M$  & $\ln \mu$ & $\hat a$ &$\hat r_0$ & $\phi_0$ & $\ln D$ & $\Delta \Omega_S$ & $\Delta \Omega_K$\\
  \hline
2 &-3.24  & -3.53 & -4.15 & -4.45 & 0.48 &-0.33 & $7.9\times 10^{-4}$ & 2.5 \\
2+3 &-3.25  &-3.54 &  -4.16 & -4.46 &-0.52 &-1.34 & $7.3\times 10^{-4}$ & $1.3\times 10^{-2}$ \\
2+3+4 &-3.25  &-3.55 &  -4.16 & -4.46 &-0.53 &-1.35 & $7.2\times 10^{-4}$ & 
$ 1.1 \times 10^{-2}$\\
  \hline
   \hline
\end{tabular}
\caption{Same as Table~\ref{tab:Schwarzschilderrnospin} but assuming a spinning 
primary with $\hat{a}=0.9$ and including $\hat a$ in the waveform parameters. In this case the SNR of the three configurations is ${\rm SNR}= 92.2,94.7, 95$, but we again normalize the errors to the fiducial value ${\rm SNR}=30$.
}\label{tab:KerreSchwarzerr}
\end{table*}

\subsection{Neglecting the spin of the secondary}\label{sec:results_nospin}

We start by neglecting the secondary spin $\chi$ from the waveform parameters. Our results are summarized in Table~\ref{tab:Schwarzschilderrnospin} and Table~\ref{tab:KerreSchwarzerr}.

Table~\ref{tab:Schwarzschilderrnospin} shows results when we also neglect the spin $\tilde a$ from the waveform parameters, and assume that both the primary and the secondary are nonspinning. In Table~\ref{tab:KerreSchwarzerr} instead, we include the spin of the primary as a parameter, injecting $\hat{a}=0.9$ but keeping all other parameters unchanged with respect to the injection of Table~\ref{tab:Schwarzschilderrnospin} (except for $\hat r_0$, since the latter changes in order for the binary to take exactly $T=1\,{\rm yr}$ to reach the ISCO). 

For $\ell=2$, our results are in very good agreement with the analysis of~\cite{Huerta:2011kt,Huerta:2011zi} which used approximated kludge waveforms. Being the latter analytical, the Fisher-matrix analysis is significantly faster than in our case. It is therefore reassuring that a fully-relativistic, numerical waveform provides the same results.

Furthermore, we find that including the octupole ($\ell=3$) contribution to the signal does not affect the measurement errors on the intrinsic parameters, but it improves the errors on the luminosity distance and on the solid angle which defines the orbital angular momentum ($\Delta \Omega_K$) by one order and two orders of magnitude, respectively. Adding the $\ell=4$ multipole does not improve such errors significantly, suggesting that $\ell>4$ multipoles are negligible for this purpose.

As expected, augmenting the dimensionality of the waveform parameter space by including the primary spin reduces the accuracy on the intrinsic parameters, especially the masses. This happens despite the fact that the ISCO frequency is higher for a rapidly-spinning BH, since we chose to normalize the results to the same SNR. For sources at a fixed distance, the SNR in the $\hat a=0.9$ case is four times larger than in the nonspinning case, almost compensating the higher dimensionality of the parameter space.

Overall, all parameters are measured with exquisite accuracy, confirming previous analyzes that used approximated semi-relativistic waveforms~\cite{Barack:2006pq,Huerta:2011kt,Huerta:2011zi,Babak:2017tow}. 
%


\subsection{Including the spin of the secondary}\label{sec:results_spin}

We now move to a more comprehensive analysis, by including the secondary spin in the waveform parameters.
We shall present two cases: with and without imposing a Gaussian prior on $\chi$. 
We start by neglecting the spin of the primary in the waveform parameters and injecting $\hat a=0$. The results of the Fisher-matrix error analysis are presented in Table~\ref{tab:Schwarzschilderrspin}, which is the extension of Table~\ref{tab:Schwarzschilderrnospin} to the case of a spinning secondary.

\begin{table*}[ht]
\centering
\begin{tabular}{cc|ccccc|ccc}
\hline
\hline
$\ell$ &prior &$\ln M$  & $\ln \mu$ &$\chi$ &$\hat r_0$ & $\phi_0$ & $\ln D$ & $\Delta \Omega_S$ & $\Delta \Omega_K$\\
  \hline
2 & no &-2.95  &-3.66 &  2.51 & -4.18& 0.55 &-0.27 & $4.4\times 10^{-3}$ & 1.6 \\
   & yes &-4.62  &-4.19 &  -0.13 & -4.96 &0.55&-0.27 & $3.1\times 10^{-3}$ & 1.5 \\
 \hline  
2+3 &no &-2.97  &-3.67 &  2.50 & -4.19 & -0.64 &-1.46 & $3.8\times 10^{-3}$  &$8.6\times 10^{-3}$\\
 &yes &-4.63  &-4.22 &-0.082 &  -4.97 & -0.66 &-1.46 & $2.4\times 10^{-3}$  &$7.9\times 10^{-3}$\\
\hline
2+3+4 &no&-2.97  &-3.67 &  2.50 & -4.19 &-0.65&-1.46 & $3.7\times 10^{-3}$ & $7.9\times 10^{-3}$ \\
&yes&-4.63  &-4.22 &-0.076&  -4.97 & -0.67 &-1.46 & $2.4\times 10^{-3}$ & $7.3\times 10^{-3}$ \\
  \hline
   \hline
\end{tabular}
\caption{Same as Table~\ref{tab:Schwarzschilderrnospin} but including a spinning secondary with $\chi=1$ and also considering the case in which a Gaussian prior on $\chi$ (with $\sigma_\chi=1$) is enforced.
}\label{tab:Schwarzschilderrspin}
\end{table*}

By comparing Table~\ref{tab:Schwarzschilderrspin} with Table~\ref{tab:Schwarzschilderrnospin} we observe some interesting features. First of all, in the case in which a prior on the secondary spin is not imposed the relative error on $\chi$ is much larger than $100\%$, confirming that this parameter is not measurable~\cite{Huerta:2011kt,Huerta:2011zi}. Nonetheless, in this case the errors on both masses deteriorate significantly (albeit they remain excellent in absolute terms).
This issue is due to nonnegligible correlations between $\chi$ and the masses.
Indeed, we find that all the intrinsic parameters are strongly correlated with $\chi$. The correlation (in absolute value) is typically $\approx 0.99$ and never less than $0.95$.
Therefore, large variations in $\chi$ as those shown in Table~\ref{tab:Schwarzschilderrspin} can correlate with a small change in the total mass or in the mass ratio.

This issue can be fixed by imposing a prior on the secondary spin, in such a way that also its errors cannot become too large. As shown in Table~\ref{tab:Schwarzschilderrspin}, imposing a Gaussian prior on $\chi$ with standard deviation $\sigma_\chi=1$ reduces the errors on this parameters, but the confidence interval is as large as the prior range, again confirming that this parameter is not measurable. (In other words, the measurement errors are dominated by the priors.)
Nonetheless, adding a prior on $\chi$ restores the accuracy in the measurements of the other intrinsic parameters, which become very similar to the case in which $\chi$ is neglected in the waveform (compare Table~\ref{tab:Schwarzschilderrspin} with prior to Table~\ref{tab:Schwarzschilderrnospin}). We also find that, including a prior on $\chi$, the correlations between $\chi$ and the other parameters are much smaller.

From Table~\ref{tab:Schwarzschilderrspin} we also observe that the role of $\ell>2$ multipoles is not affected by the secondary spin: also in this case the inclusion of the $\ell=3$ multipole improves the errors on the distance and on the orbital angular momentum solid angle by one and two orders of magnitude, respectively.

Finally, we are now in a position to present the complete analysis by including both the spin of the primary and of the secondary.
A summary of our results are presented in Table~\ref{tab:Kerrerrspin0.9-0.99}
for the cases with $\hat a =0.9$ and $\hat a = 0.99$, and considering both $\mu=10 M_\odot$ and $\mu=100 M_\odot$. In this analysis we only include the quadrupole ($\ell=2$) since anyway the higher multipoles do not affect the errors on the intrinsic parameters.

\begin{table*}[ht]
\centering
\begin{tabular}{ccc|cccccc}
\hline
\hline
$\tilde a_{\rm injected}$ &
$\mu/M_\odot$ & prior & $\ln M$  & $\ln \mu$ & $\hat a$ & $\chi$ &$\hat r_0$ & $\phi_0$\\
  \hline
  \multirow{4}{*}{0.9}&
\multirow{2}{*}{10} &no &-2.26  &-2.41 &  -2.66 & 2.85 & -3.88 & 0.48 \\
& & yes &-3.24  &-3.53 &  -4.14 & 0.48 & -4.45 &0.48\\
& \multirow{2}{*}{100} & no &-2.20  &-2.39 &  -2.78 & 1.66 & -4.14 &-0.015\\ 
& & yes &-3.30  &-3.52 & -4.32 & 0.064 & -4.93 & -0.024\\
\hline
\multirow{4}{*}{0.99} &
\multirow{2}{*}{10} &  no & -2.81  &-2.96 &  -4.55 & 1.98 & -3.89 & 0.47\ \\
& & yes & -3.51  &-3.76 &  -4.67 & 0.52 & -4.32 & 0.47\\
& \multirow{2}{*}{100} & no& -2.14  &-2.33 &  -3.39 & 1.21 & -3.75 & -0.12  \\
& & yes& -3.01  &-3.22 &  -4.03 & 0.11 & -4.50 & -0.12  \\
\hline
\hline
\end{tabular}
\caption{Fisher-matrix errors on the EMRI parameters including both binary components spin in the waveform and including a spinning secondary with $\chi=1$. We include only the quadrupole ($\ell=2$) in the signal and consider two choices of the mass ratios and two values of the primary spin, with and without imposing a Gaussian prior on $\chi$. 
In these configurations, the ${\rm SNR}$ for $\mu = 10 M_\odot (100 M_\odot$) is ${\rm SNR}=92.2$ (${\rm SNR}=174$) when $\hat a=0.9$ and ${\rm SNR}=100$ (${\rm SNR}=195$) when $\hat a=0.99$. However, also in this table the results have been rescaled to have 
${\rm SNR}=30$ (${\rm SNR}=150$) when $\mu = 10 M_\odot (100 M_\odot$), regardless of the primary spin.
}
\label{tab:Kerrerrspin0.9-0.99}
\end{table*}

Also in this general case we observe the same features of the previous analyses. In particular, the secondary spin is not measurable but its inclusion can significantly deteriorate the accuracy in the measurements of the masses, unless a prior on $\chi$ is enforced.
Even in an extreme case ($\hat a=0.99$, $\mu=100M_\odot$) the relative error on $\chi$ is larger than $100\%$ for ${\rm SNR}<2433$. 
Also in this general case, we find that including the secondary spin with a prior yields the same errors as in the case in which $\chi$ is neglected in the waveform parameters.

\section{Conclusion}\label{sec:conclusion}
EMRIs are unique GW sources that can be potentially used to tests fundamental physics and astrophysics to unprecedented levels.
However, this huge potential comes with its own burden: data analysis and parameter estimation of EMRIs are challenging and, in many respects, still an open issue.

In this work we have focused on circular equatorial motion around a Kerr BH and computed the waveform numerically to leading order in an adiabatic expansion, taking into account the motion of the LISA constellation, higher harmonics, and also including the leading correction from the spin of the secondary in the post-adiabatic approximation. We have then performed a brute-force Fisher-matrix analysis without resorting to approximated or kludge waveforms. Clearly our approach is very time-consuming and inefficient for practical purposes, but can be used to quantify the accuracy of approximated waveforms that are instead much more efficient for EMRI parameter estimation. 
Our analysis confirmed that using approximated (and dramatically more efficient) waveforms~\cite{Huerta:2011kt,Huerta:2011zi,Babak:2017tow} does not significantly affect the measurement errors on the binary's parameters, including the subleading spin of the secondary.

The measurability of the secondary spin is particularly interesting for various applications, including model-agnostic tests of the Kerr hypothesis~\cite{Piovano:2020ooe,Piovano:2020zin}. We have therefore performed a detailed analysis on this aspect. We confirm the results of Refs.~\cite{Huerta:2011kt,Huerta:2011zi} which, using approximated waveforms, found that the secondary spin is not measurable, although it produces a nonnegligible dephasing~\cite{Piovano:2020ooe,Piovano:2020zin}. This is due to correlations that exist between the secondary spin and the other intrinsic parameters. 
Because of these correlations, even if the secondary spin is not measurable, its inclusion in the waveform model can deteriorate the accuracy on the measurements of other parameters by orders of magnitude, unless a physically-motivated prior on the secondary spin is imposed. In the latter case, we find that the Fisher-matrix errors are identical to those obtained neglecting the secondary spin in the waveform parameters. This further suggests that, for EMRIs, the secondary spin is negligible for parameter estimation.

Finally, we found that including higher harmonics in the GW signal improves the errors on the luminosity distance by an order of magnitude and those on the binary orbital angular-momentum angles by two orders of magnitude, relative to the quadrupole-only case.
This is particularly relevant to identify the environment where EMRIs form~\cite{AmaroSeoane:2007aw,Pan:2021oob}, for possible applications of multimessenger astronomy with EMRIs~\cite{McGee:2018qwb} and for prospects to use EMRIs as standard sirens~\cite{Laghi:2021pqk}.

Our brute force analysis should be intended as a proof-of-concept aimed at assessing the accuracy of more efficient (but approximated) methods which, after a positive benchmark, can be used more confidently in parameter estimation.
At the same time our analysis can and should be extended in various directions, to provide a necessary benchmark for more complete waveforms, for example the recent ones obtained by using order-reduction and deep-learning techniques for eccentric nonspinning orbits around Schwarzschild~\cite{Katz:2021yft,Chua:2020stf}. Obvious extensions of our work are the inclusion of eccentricity and nonequatorial orbits, as well as spin misalignment. Finally, our waveform does not include all the next-to-leading order terms in an adiabatic expansion, in particular it lacks the leading-order conservative self-force corrections. Including all these interesting effects is left for future work.

\begin{acknowledgments}
This work makes use of the Black Hole Perturbation Toolkit and \textsc{xAct} \textsc{Mathematica} package. 
Numerical computations were performed at the Vera cluster of the Amaldi Research Center funded by the MIUR program  ``Dipartimento di 
Eccellenza" (CUP:~B81I18001170001).
P.P. and R.B. acknowledge financial support provided under the European Union's H2020 ERC, Starting 
Grant agreement no.~DarkGRA--757480. We also acknowledge support under the MIUR PRIN and FARE programmes (GW-NEXT, 
CUP:~B84I20000100001), and networking support by the COST Action CA16104.
G.A.P. would like to thank Luca Graziani for the support during the computational runs on Vera cluster of the Amaldi Research Center.
\end{acknowledgments}

\appendix

\section{Teukolsky equation in hyperboloidal-slicing coordinates}\label{sec:app HST eq}

The coefficients $\tilde F(\hat r)$ and $\tilde U(\hat r)$ of Eq.~\eqref{eq:radialHST} are given by
\begin{align}
\tilde F(\hat r;H)& = \frac{2}{\hat r^2 +\hat a^2} \bigg( \hat r^2 -\hat a^2 -\tilde G(\hat r;H) \bigg) \, , \\ 
\tilde G(\hat r;H) &= (\hat r^2 +\hat a^2)[s(\hat r -1)- i ( (\hat r^2 +\hat a^2)\hat \omega H +m \hat a)] +\nonumber \\
 & +\frac{\hat a^2 \Delta}{\hat r} \, , \\
\tilde U(\hat r;H) &= 2is\hat \omega [\hat r \Delta (1-H)-(\hat r^2 -\hat a^2)(1+H)]+ \nonumber \\\
&+\frac{\Delta}{\hat r^2} [2\hat a^2-\hat r^2\lambda_{\ell m \hat \omega}-2\hat r (s+1)] +\nonumber \\\
& -2m\hat a \hat \omega (\hat r^2 +\hat a^2)(1+H)-2i \hat a \frac{\Delta}{\hat r} (m+\hat a \hat \omega H) \, ,
\end{align}
where $H = -1 \, (+1)$ for the linearly independent solution $\psi^{\textup{in}} (\psi^{\textup{up}})$. This is the same convention adopted in the \textit{Teukolsky} package of the Black Hole Perturbation Toolkit~\cite{BHPToolkit}. Notice that
\begin{align}
\tilde U(\hat{r}_+;-1)& = 0 \ , \\
\frac{\tilde U(\hat{r}\to \infty;1)}{\Delta^2}  &\to -\frac{\lambda_{\ell m \hat \omega}+4 a m \hat \omega + 4 i s \hat \omega}{\hat r^2} \,, \\
\frac{\tilde F(\hat{r}\to \infty;1)}{\Delta} &\to 2i\hat \omega \,.
\end{align} 
It is easy to show that the ordinary differential equation~\eqref{eq:radialHST} has three singularities on the real positive axis: two at the horizons $\hat{r}=\hat{r}_-$  and $\hat{r}=\hat{r}_+$, both 
of which are regular singularities, and one at $\hat{r}=\infty$ which is an irregular singularity of rank $1$. Despite having different coefficients, the radial Teukolsky equation, the Sasaki-Nakamura equation, and Eq.~\eqref{eq:radialHST} have the same singularities. Therefore, both the Sasaki-Nakamura transformation and transformation~\eqref{eq:ansatzHST} preserve the singularity structure of the radial Teukolsky equation.
We compute accurate boundary conditions at the outer horizon $\hat r_+$ and at infinity through suitable series expansions, as done in Ref.~\cite{Piovano:2020zin}. The Fuchs theorem guarantees that the solutions of~\eqref{eq:radialHST} around $\hat{r}_+$ can be written as Frobenius series, with radius of convergence
\begin{equation}
\hat{r}_+ - \hat{r}_ -= 2 \sqrt{1-\hat{a}^2} \ .
\end{equation}
At infinity or when $\hat{a}=1$ (for which $\hat{r}_+=\hat{r}_-$), the boundary conditions 
can be computed accurately as asymptotic expansions.

\subsection{Boundary conditions for the Teukolsky equation in hyperboloidal-slicing coordinates}\label{sec:HSC BCs}
\subsubsection{Boundary condition at the horizon}
To compute the boundary conditions at the outer horizon $\hat{r}_+$, it is 
convenient to rewrite Eq.~\eqref{eq:radialHST} as
\begin{equation}
(\hat{r}-\hat{r}_+)^2\frac{d^2 \psi^{\textup{in}}}{d\hat{r}^2} + (\hat{r}-\hat{r}_+) p_H(\hat{r})\frac{d\psi^{\textup{in}}}{d\hat{r}} + q_H(\hat{r})\psi^{\textup{in}}=0 \, ,
\end{equation}
where 
\begin{align}
p_H(\hat{r}) &= \frac{\tilde F(\hat r;-1)}{\hat r -\hat r_-}\,, \qquad  q_H(\hat{r}) = \frac{\tilde U(\hat{r};-1)}{(\hat r - \hat r_-)^2} \ .
\end{align}
We seek for a Frobenius power series solution of the form
\begin{equation}
\psi^{\textup{in}} = (\hat{r}-\hat{r}_+)^d \displaystyle\sum_{n=0}^{\infty} a_n (\hat{r}-\hat{r}_+)^n \ , \label{eq:horexp}
\end{equation}
where the index $d$ is a solution of the indicial equation
\begin{equation}
I(d) = d(d-1) + p_H(\hat{r}_+) d + q_H(\hat{r}_+) =0\ .
\end{equation}
For Eq.~\eqref{eq:radialHST}, the latter is given by
\begin{equation}
I(d) = d(d -c_H)= 0 \ , \quad \ c_H = \frac{4i\hat r_+}{\hat r_+-\hat r_-}\kappa + s \ ,
\end{equation}
and $\kappa =\hat{\omega} -m \hat a/(2 \hat{r}_+)$. Near the outer horizon $\hat r_+$, the radial solution $R^{\textup{in}}_{\ell m \hat \omega}$ has the following asymptotic behavior
\begin{align}
\Rin &\sim \Delta^{-s} e^{-i \hat{\kappa} \hat{r}^\ast} \qquad  \hat{r} \to \hat r_+  \, ,
\end{align}
Thus, only $d=0$ is a physical solution of the indicial equation. Moreover, we notice that the ansatz~\eqref{eq:ansatzHST}  for the $R^{\textup{in}}_{\ell m \hat \omega}$ solution can be rewritten as
\begin{align}
\Rin(\hat r) &= \hat r^{-1} \Delta^{-s} e^{- i \kappa \hat r^*}e^{-i \delta_H(\hat r)} \psi^{\textup{in}}(\hat r) \ , \\
\delta_H(\hat r) &\equiv \frac{am}{\hat r_+} \Big[\frac{\hat r}{2} +\ln \Big(\frac{\hat r -\hat r_-}{2}\Big)\Big] \,.
\end{align}
Therefore, to ensure the correct physical behavior of $\Rin(\hat r)$ at the outer horizon, we fix $d=0$ and write the Frobenius series~\eqref{eq:horexp} as
\begin{equation}
\psi^{\textup{in}}= \hat r_+ e^{i \delta_H(\hat r_+)}\displaystyle \sum_{n=0}^{\infty} 
a_n (\hat r-\hat r_+)^n \label{eq:BCHor} \ .
\end{equation}
The recursion relation for the coefficients $a_n$ is (setting $a_0=1$)
\begin{equation}
a_n = - \frac{1}{I(n)}  \displaystyle\sum_{k=0}^{n-1}\Big(k\, p_H^{(n-k)}(\hat{r}_+) +q_H^{(n-k)}(\hat r_+)\Big) a_k 
\ ,
\end{equation}
where $p_H^{(k)}(\hat{r}_+) $ and $q_H^{(k)}(\hat{r}_+) $ are the $k$-th 
derivatives of the coefficients $p_H(\hat{r})$ and $q_H(\hat{r})$ with respect to $\hat{r}$, and calculated at $\hat{r}_+$. Their general expression is given by
\begin{widetext}
\begin{align}
p_H^{(n)}(\hat{r}_+) =
\begin{cases}
1 -c_H \quad  & n=0 \ , \\
(\rho_H^2\hat r_+)^{-1}[-2\hat r^2_- +\hat a^2 (3+2 s +4 i \hat r_+ \hat \omega)+ \hat r_+ (-2 i\hat a m +2i \hat a^2 \hat \omega - (\hat r_+ +2s +2 i \hat r_+^2 \hat \omega ))] \quad & n=1  \ , \\
2(-\hat r_+)^{-n} -\rho_H^{-n}+\rho_H^{-n-1}[2s\hat r_- +2i \hat r_-^2 \hat \omega + 2 i(-\hat a m + is +\hat a^2 \hat \omega)] \quad  & n>1 \ , \\
\end{cases}
\end{align}
\begin{align}
q_H^{(n)}(\hat{r}_+) =
\begin{cases}
0 \quad  & n=0 \ , \\
(\rho_H\hat r_+)^{-1}[2 i \hat a m  +2 (s-1) - 2i \hat a^2 \hat \omega + \hat r_+ (2+\lambda_{\ell m \hat \omega}-4 i \hat r_+ s\hat \omega)] \quad & n=1  \ , \\
2(n-1)(-\hat r_+)^{-n}+ \rho_H^{-n}\Big[(2+\lambda_{\ell m \hat \omega}-4 i \hat r_- s \hat \omega)+\frac{2 n}{\hat r_+}(s-1+i \hat a (m -\hat a \hat \omega)) \!\prescript{}{2}{F_1}\Big(1,1-n;2;\frac{\hat r_-}{\hat r_+}\Big)\Big] \quad  & n>1 \ , \\
\end{cases}
\end{align}
\end{widetext}
where $\rho_H \equiv (\hat r_- - \hat r_+)$ and $\!\prescript{}{2}{F_1}(1,1-n;2; \hat r_-/\hat r_+)$ is the hypergeometric function $\!\prescript{}{2}{F_1}(a,b;c; z$) .
\subsubsection{Boundary condition at infinity}
General expressions for series solutions around irregular singularities are also available in the literature~\cite{Olver:1994:AEC,Olver:1997:ASL,Olver:1974asymptotics}. However, unlike the regular case, these solutions are not convergent, and  have to be considered as asymptotic expansions.
To calculate the boundary conditions at infinity, we rewrite Eq.~\eqref{eq:radialHST} as
\begin{equation}
\frac{d^2\psi^{\textup{up}}}{d\hat{r}^2} + p_\infty(\hat r)\frac{d\psi^{\textup{up}}}{d\hat{r}} + 
q_\infty(\hat r) \psi^{\textup{up}}=0\ ,
\end{equation}
where 
\begin{align}
p_\infty(\hat r) &= \frac{\tilde F(\hat r;1)}{\Delta}\,, \qquad  q_\infty(\hat r) = \frac{\tilde U(\hat{r};1)}{\Delta^2} \ .
\end{align}
The functions $p_\infty(\hat{r})$ and $q_\infty(\hat{r})$ are analytic on 
the positive real axis, so the series
\begin{alignat*}{2}
p_\infty(\hat{r}) &= \displaystyle \sum_{n=0}^{\infty} \frac{1}{n!}  \frac{p_\infty^{(n)}}{\hat{r}^n} \ ,  \qquad 
q_\infty(\hat{r}) &= 
\displaystyle \sum_{n=0}^{\infty} \frac{1}{n!}  \frac{q_\infty^{(n)}}{\hat{r}^n} \ ,
\end{alignat*}
converge, with $p_\infty^{(n)}$ and $q_\infty^{(n)}$ being the $n$-th derivatives of 
the coefficients $p_\infty$ and $q_\infty$  with respect to $\hat{r}$. In the case of irregular singularities of rank 1, the formal solution is given by 
\begin{equation}
\psi^{\textup{up}} = e^{\gamma \hat{r}} \hat{r}^\xi \displaystyle\sum_{n=0}^{\infty} \frac{b_n}{\hat{r}^n} \ ,
\end{equation}
provided that at least one of $p_\infty^{(0)}$, $q_\infty^{(0)}$ or $q_\infty^{(1)}$ is 
nonzero. The exponent $\gamma$ is one of the solutions of the characteristic equation
\begin{equation}
\gamma^2 + p^{(0)}_\infty \gamma + q^{(0)}_\infty =0 \ ,
\end{equation}
while
\begin{equation}
\xi= -\frac{p_\infty^{(1)}\gamma +q_\infty^{(1)}}{p_\infty^{(0)}+2 \gamma} \ .
\end{equation}
For Eq.~\eqref{eq:radialHST} we have:
\begin{align}
q_\infty^{(0)} &= 0= q_\infty^{(1)} \ , \qquad p_\infty^{(0)} = 2i \hat\omega \ , \qquad  p_\infty^{(1)} = 4 i \hat{\omega}-2 s \ 
,\\
\gamma( \gamma&+ 2i\hat\omega) =0  \ ,\qquad \xi = -\frac{\gamma (2i\hat \omega -s)}{\gamma+ i \hat \omega} \ .
\end{align}
When $\hat{r} \to \infty$, the radial solution $R^{\textup{up}}_{\ell m \hat \omega}$ has the following asymptotic behavior
\begin{align}
\Rup &\sim r^{-(2s+1)} e^{i \hat \omega \hat{r}^\ast} \qquad  \hat{r} \to \infty  \, .
\end{align}
Thus, only $\gamma=0$ is a physical solution of the characteristic equation, and we can write
\begin{equation}
\psi^{\textup{up}}= \displaystyle 
\sum_{n=0}^{\infty}\frac{b_n}{\hat{r}^n} \,.
\label{eq:BCInf}
\end{equation}
The general recursion relation for the coefficients $b_n$ is (we set again 
$b_0=1$):
\begin{align}
(p_\infty^{(0)} + 2 \gamma) n b_n = (n - \xi)(n-1 -\xi)b_{n-1} +  \nonumber\\
+\displaystyle \sum_{k=1}^{n}\Big[\gamma p_\infty^{(k+1)}+q_\infty^{(k+1)} - (n-k-\xi)p_\infty^{(k)}\Big]b_{n-k}\ .
\end{align}
In our case, we can write
\begin{equation}
b_n = \frac{n-1}{2i\hat \omega}b_{n-1} + \frac{1}{2i\hat \omega n}  \displaystyle\sum_{k=1}^{n}\Big[q_\infty^{(k+1)}- (n-k)p_\infty^{(k)}\Big]b_{n-k} 
\ ,
\end{equation}
where
\begin{align}
p_\infty^{(n)} & =
\begin{cases}
2 i \hat \omega\quad  & n=0 \ , \\
4i \hat \omega -2s \quad & n=1  \ , \\
\hat r_-^{n-1}+\hat r_+^{n-1} + P_- -P_+ \quad  & n>1 \ , \\
\end{cases} \\
P_\pm &= \frac{2 \hat r_\pm^{n-1}}{\rho_H}[(1-\hat r_\pm)s + i(\hat a m +(\hat r^2_\pm +\hat a^2)\hat \omega)]\ ,
\end{align}
and
\begin{align}
q_\infty^{(n)} =
\begin{cases}
0 \quad  & n=0,1 \ , \\
-(4\hat a m \hat \omega + 4 i s \hat \omega +\lambda_{\ell m \hat \omega}  ) \quad & n=2  \ , \\
\frac{2}{\rho_H}Q_1 + \frac{4 \hat \omega}{\rho_H^3}Q_2 \quad  & n>2 \ , \\
\end{cases}
\end{align}
with
\begin{align}
Q_1 &=\hat r_-^{n-2}\hat r_+ - \hat r_- \hat r_+^{n-2} -\frac{1}{2}(\hat r_-^{n-1}- \hat r_+^{n-1})\lambda_{\ell m \hat \omega} +\nonumber \\
 &- (i\hat a m+s+1+i\hat a^2 \hat \omega) (\hat r_-^{n-2}-\hat r_+^{n-2}) \ , \\[0.5em]
Q_2&=i s \hat a^2[ \rho_H(n-1) (\hat r_-^{n-2}+\hat r_+^{n-2}) -2(\hat r_-^{n-1} -\hat r_+^{n-1} ) ]+ \nonumber \\
&+(i s+\hat a m) [\hat r_-^n(2 - n\rho_H)-\hat r_+^n(2 + n\rho_H)] + \nonumber \\
&+ \hat a^3 m [ \rho_H(1-n) (\hat r_-^{n-2}+\hat r_+^{n-2}) +2(\hat r_-^{n-1} -\hat r_+^{n-1} ) ] + \nonumber\\
&-\frac{i}{2}\rho^2_H \hat a^2 (\hat r_-^{n-2}-\hat r_+^{n-2})\ .
\end{align}

\section{Linearization in the secondary spin}\label{sec:appe_lin}
\subsection{Linearization of the angular Teukolsky equation}\label{sec:lin angular Teu}
For the study of the eigenvalues and eigenfunctions of Eq.~\eqref{eq:swsweq}, it is convenient to perform a change of variable defining $x = \cos \theta$, obtaining 
\begin{align}
\mathcal{H} | S \rangle &= - \lambda_{\ell m\hat\omega}|S \rangle\,,
\quad  | S \rangle \equiv \rswsh\,,  
\quad \mathcal{H} = \mathcal K + \mathcal V   \,,
\end{align}
with
\begin{align}
\mathcal{K} & \equiv \totder{}{x}\bigg((1-x^2)\totder{}{x} \bigg) \,, \\
\mathcal{V}& \equiv c x (c x -2s) -c^2 +s+ 2mc  - \frac{(m +s x)^2}{1-x^2}  \,,\label{eq:swsweq2}
\end{align}
where the dependence on the spin perturbation $s$ is understood to reduce clutter in the notation. We consider here only the case in which $c \in \mathbb{R}$. Physical solutions of~\eqref{eq:swsweq2} must be regular in the interval $[-1,1]$, which entails that $\ell$ and $m$ must be integers with $ |m| \leq \ell $. The solutions to Eq.~\eqref{eq:swsweq2} can be written as a series expansion around the singular points $x = \pm 1$ ~\cite{Leaver:1985ax,Leaver1986SolutionsTA}:
\begin{equation}
S^c_{\ell m} =\frac{e^{c x}}{\sqrt{\mathcal{N}}}(1+x)^{k_-}(1-x)^{k_+} \displaystyle\sum_{n=0}^{\infty} d_n (1+x)^n \ , \label{eq:Leaverseries}
\end{equation}
where $k_\pm = |m \pm 2|/2$ and the coefficients $d_n$ are given by the three-term recursion relations
\begin{align}
&\alpha_0 d_1 + \beta_0 d_0 = 0 \,, \\
& \alpha_n d_{n+1} +\beta_n d_n + \gamma_n d_{n-1} = 0  \qquad n=1,2 \dots 
\end{align}
with
\begin{align}
\alpha_n &= - 2(n+1)(n+2 k_-+1 ) \,, \\
\beta_n &= n(n+1)+ 2 n(k_s + 1-2c) - 2 c(2 k_-+s+1) +\nonumber \\
&+ k_s(k_s+1) -s(s+1)-\lambda_{\ell m \hat \omega} -2m c \,, \\
\gamma_n &= 2 c (n +k_s + s) \,,
\end{align}
and $ k_s = k_+ + k_-$. The normalization constant $\mathcal N$ can be written analytically as
\begin{equation}
\mathcal N \equiv \int_{-1}^1 \!\!(S_{\ell m}(x))^2\dd x = (2\pi) 2^{1+2k_s}e^{-2c}\Gamma(1+2k_+)\mathfrak{N} \,,
\end{equation}
where
\begin{align}
&\mathfrak N \equiv  \displaystyle\sum_{n=0}^{\infty}\frac{\Gamma(1+2k_- +n) }{\Gamma(2+2k_s+n)}2^n F(n,n;c)\displaystyle\sum_{i=0}^n d_i d_{n-i} \,,\\
&F(n,n;c) \coloneqq \prescript{}{1}{F_1}(1+2k_-+n, 2+2 k_s+n;4c)\,,
\end{align}
while $\Gamma(z)$ is the Euler gamma function and $\prescript{}{1}{F_1}(a, b;z)$ is the Kummer confluent hypergeometric function.
To ensure the convergence of the series~\eqref{eq:Leaverseries} at $x = \pm 1$, the eigenvalue $\lambda_{\ell m\hat\omega}$ must satisfy the implicit continued fraction
\begin{equation}
0 = \beta_0 - \frac{\alpha_0 \gamma_1}{\beta_1-}\frac{\alpha_1\gamma_2}{\beta_2-}\frac{\alpha_2\gamma_3}{\beta_3-}\dots
\end{equation}
With the requirement of regularity at the boundaries $[-1,1]$, Eq.~\eqref{eq:swsweq2} defines a Sturm-Liouville eigenvalue problem. In particular, the eigenvalue problem is singular because the coefficient $(1-x^2)$ vanishes at the boundaries. Nevertheless, it can be shown that Eq.~\eqref{eq:swsweq2} still satisfies many of the properties of a regular Sturm-Liouville problem, namely (see~\cite{Borissov:2009bj} and references therein):
\begin{itemize}
\item the operator $\mathcal H$ is Hermitian, i.e. $\langle v| \mathcal{H} | w \rangle = \langle w| \mathcal{H} | v \rangle $ for any vector $v, w$;
\item given a set $s,m,c$, the functions $\rswsh(\theta)$ form a (strong) complete, orthogonal set on $[-1,1]$, labeled by the additional integer $\ell$ (see~\cite{Stewart:1975});
\item each eigenvalue  $\lambda_{\ell m \hat \omega}$ has (up to a constant) a unique eigenfunction for any set $s,m,c$.
\end{itemize}
Thus, we can conveniently treat the secondary spin $\sigma$ as a small perturbation of an Hermitian operator and compute the linear corrections in $\sigma$ to $\lambda_{\ell m \hat \omega}$ using the same techniques of nondegenerate perturbations of a quantum mechanical system~\cite{Sakurai:quantum_mechanics}. To linear order in $\sigma$, we obtain
\begin{align}
&\mathcal{H}^0 | S^0 \rangle = -\lambda^0_{\ell m} | S^0 \rangle  \, ,  \\
&\mathcal{H}^0| S^1 \rangle +\mathcal{V}^1| S^0 \rangle = -\lambda^0_{\ell m} | S^1 \rangle -\lambda^1_{\ell m} | S^0 \rangle \, , \\
& \mathcal H^0 = \mathcal{K} +\mathcal{V}^0 \, ,  \\
&\mathcal V^1 =2c^1(c^0x^2 - s x+m-c^0) \, , 
\end{align}
where $\mathcal{V}^0$ is simply given by $\mathcal{H}$ with $c \leftrightarrow c^0$, $S^0_{\ell m}\equiv | S^0 \rangle, S^1_{\ell m} \equiv | S^1 \rangle $ and
\begin{widetext}
\begin{align}
\lambda^1_{\ell m} &=\langle S^0 |\mathcal{V}^1|S^0 \rangle \equiv \int_{-1}^1 \!\!S^0_{\ell m}\mathcal{V}^1S^0_{\ell m}\dd x 
 = -\frac{c^1}{\mathfrak{N}^0}\displaystyle\sum_{n=0}^{\infty}\Xi(n)\Big[\Upsilon(n) F(n,n+1;c^0)-\Pi(n)F(n,n;c^0)\Big]\displaystyle\sum_{i=0}^n d^0_i d^0_{n-i} \,,
\end{align}
\end{widetext}
with
\begin{align}
\Xi(n) & \equiv 2^{n+1} \frac{\Gamma(1+2k_-+n)}{\Gamma(3+k_s+n)} \, , \\
\Upsilon(n) & \equiv  (1+2k_+)(2+2k_s+n+2s) \, , \\
\Pi(n) &\equiv (2+2k_s+n)(1+2k_+-m+s) \, .
\end{align}
The term $\mathfrak N^0$ is given by $\mathfrak N$ with $c \leftrightarrow c^0$. We computed the 0th order eigenvalue $\lambda^0_{\ell m}$, the corresponding eigenfunctions $S^0_{\ell m}$ and the coefficients $d^0_n$ using the routines of the \textit{SpinWeightedSpheroidalHarmonics} \textsc{Mathematica} package of~\cite{BHPToolkit}.
Once the correction to the eigenvalue $\lambda^1_{\ell m}$ is known, we can evaluate the correction to the eigenfunction $S^1_{\ell m}$ by expanding in $\sigma$ the Leaver series~\eqref{eq:Leaverseries}, obtaining
\begin{align}
S^1_{\ell m} &=\frac{e^{c^0 x}}{\sqrt{\mathcal N^0}}(1+x)^{k_-}(1-x)^{k_+} \displaystyle\sum_{n=0}^{\infty}\Big[d^1_n (1+x)^n + \nonumber \\
& +d^0_n(1+x)^n\Big(\!(1+x)-\frac{\mathfrak N^1}{2 \mathfrak N^0}\Big)\Big]  ,
\end{align}
where the three-term recursion relation for the correction $d^1_n$ is given by, for $n=1,2 \dots$
\begin{align}
& d^1_0 = 0  \qquad    \alpha_0 d^1_1+ \beta^1_0 d^0_0 =0 \,, \\
&\alpha_n d^1_{n+1}+ \beta^0_n d^1_n+\beta^1_n d^0_ n+\gamma^0_n d^1_{n-1} + \gamma^1_n d^1_{n-1} =0   \ ,
\end{align}
with
\begin{align}
\beta^1_n &= -2c^1 (1+2k_- +m + 2 n +s) -\lambda^1_{\ell m}  \,, \\
\gamma^1_n &= 2 c^1 ( k_s + s +n) \,,
\end{align}
and
\begin{align}
&\mathfrak N^1 \equiv  \displaystyle\sum_{n=0}^{\infty} \frac{2^{n+1}\Gamma(1+2k_+ +n) }{\Gamma(2+2k_s+n)}\Big[F(n,n;c^0)\displaystyle\sum_{i=0}^n d^0_i d^1_{n-i} + \nonumber\\
&+2\frac{1+2k_-+n}{2+2k_s+n}F(n+1,n+1;c^0)\displaystyle\sum_{i=0}^n d^0_i d^0_{n-i} \Big]\,.
\end{align}

\subsection{Linearization of the radial Teukolsky equation}\label{sec:lin radial Teu}
The linear corrections in $\sigma$, $R^{\textup{in},1}_{\ell m}$ and $R^{\textup{up},1}_{\ell m}$, were obtained by expanding the ansatz~\eqref{eq:ansatzHST} as follows. Let us first define 
\begin{align}
N^0_\mp &= \hat r^{-1} \Delta^{-s} e^{\mp i \hat \omega^0 \hat r^*}e^{i m \tilde \phi} \ , \\
D^0_\mp &=-\frac{N^0_\mp}{\Delta}  \Big( \frac{\Delta}{\hat r} + 2 s (\hat r -1)\pm i (\hat r^2 +\hat a^2)\hat \omega^0 + i\hat a m \Big)\ ,\\
D^1_\mp &= \mp i \omega^1  \Big(\frac{\hat r^2+\hat a^2}{\Delta}N^0_\mp+ \hat r^\ast D^0_\mp\Big) \ ,
\end{align}
It is possible then to write
\begin{align}
R^{\alpha,0}_{\ell m} &= N^0_\mp  \psi^{\alpha,0}\ , \\
R^{\alpha,1}_{\ell m} &= N^0_\mp ( \psi^{\alpha,1} \mp i \hat \omega^1 \hat r^* \psi^{\alpha,0} )\ ,\\
\totder{R^{\alpha,0}_{\ell m}}{\hat r} &=  \psi^{\alpha,0} D^0_\mp + N^0_\mp \totder{\psi^{\alpha,0}}{\hat r} \, ,\\
\totder{R^{\alpha,1}_{\ell m}}{\hat r} &=\psi^{\alpha,1} D^0_\mp +  \psi^{\alpha,0}D^1_\mp+\\
&+ N^0_\mp\Big(\totder{\psi^{\alpha,1}}{\hat r} \mp i \hat \omega^1 \hat r^* \totder{\psi^{\alpha,0}}{\hat r} \Big)\ ,
\end{align}
where $\alpha = \text{in}\,(\text{up})$ for the minus (plus) sign. 
Finally, we computed the linear corrections $\psi^{\textup{in},0},\psi^{\textup{in},1}$ and $\psi^{\textup{up},0},\psi^{\textup{up},1}$ as solutions of a system of ordinary differential equations obtained by expanding Eq.~\eqref{eq:radialHST} and the related boundary conditions in $\sigma$. 

For the solutions  $\psi^{\textup{in},0},\psi^{\textup{in},1}$, the system of differential equations is
\begin{align}
&\frac{d^2 \psi^{\textup{in},0}}{d\hat{r}^2} + \frac{p^0_H(\hat r)}{\hat r -\hat r_+}\frac{d\psi^{\textup{in},0}}{d\hat r} + \frac{q^0_H(\hat r)}{(\hat r -\hat r_+)^2}\psi^{\textup{in},0}=0 \, , \\
&\frac{d^2 \psi^{\textup{in},1}}{d\hat{r}^2} + \frac{p^0_H(\hat r)}{\hat r -\hat r_+}\frac{d\psi^{\textup{in},1}}{d\hat r} + \frac{p^1_H(\hat r)}{\hat r -\hat r_+}\frac{d\psi^{\textup{in},0}}{d\hat r} + \nonumber \\
&+ \frac{q^0_H(\hat r)}{(\hat r -\hat r_+)^2}\psi^{\textup{in},1} + \frac{q^1_H(\hat r)}{(\hat r -\hat r_+)^2}\psi^{\textup{in},0}=0 \, ,
\end{align}
where
\begin{align}
&p^1_H(\hat{r}) = -\frac{2\tilde G^1(\hat r;-1)}{(\hat r -\hat r_-)(\hat r^2 +\hat a^2)} \qquad  q^1_H(\hat{r}) = \frac{\tilde U^1(\hat{r};-1)}{(\hat r - \hat r_-)^2} \,, \\
&\tilde G^1(\hat r;-1) = i (\hat r^2 +\hat a^2)^2 \hat \omega^1 \,, \\
&\tilde U^1(\hat r;-1) = \Delta \Big[-\lambda^1_{\ell m} + 2i\hat \omega^1\Big(\frac{\hat a^2}{\hat r}  + 2\hat r  s \Big)\Big]
\end{align}
and the boundary conditions for $\psi^{\textup{in},1}$ are
\begin{equation}
\psi^{\textup{in},1}(\hat r)= \hat r_+ e^{i \delta_H(\hat r_+)}\displaystyle \sum_{n=0}^{\infty} 
a^1_n (\hat r- \hat r_+)^n \ .
\end{equation}
The recursion relation for the coefficients $a^1_n$ is (setting $a^1_0=0$)
\begin{align}
a^1_n  &=  -\displaystyle\sum_{k=0}^{n-1}\Big(k p_H^{(n-k),1}(\hat r _+) +q_H^{(n-k),1}(\hat r_+)\Big)\frac{a^0_k}{I(n)} + \nonumber \\
& - \displaystyle\sum_{k=0}^{n-1}\Big(k p_H^{(n-k),0}(\hat r_+) +q_H^{(n-k),0}(\hat r_+)\Big)\frac{a^1_k}{I(n)} - \frac{c^1_H a^0_n}{n- c^0_H}  
\end{align}
where $c^1_H = \frac{4i\hat r_+}{\hat r_+-\hat r_-}\hat \omega^1$ and
\begin{align}
p_H^{(n),1}(\hat{r}_+) =
\begin{cases}
 -c^1_H \quad  & n=0 \ , \\[0.2em]
-2 i (\hat r_+^2 - 3 \hat a^2)\hat \omega^1\rho_H^{-2} \quad & n=1  \ , \\[0.2em]
2i (\hat a^2 +\hat r^2_-)\rho_H^{-1-n}\hat \omega^1 \quad  & n>1 \ , 
\end{cases}
\end{align}
\begin{align}
q_H^{(n),1}(\hat{r}_+) =
\begin{cases}
0 \quad  & n=0 \ , \\
\frac{\hat r_+(\lambda^1_{\ell m}-4 i \hat r_+ s \hat \omega^1)-2 i \hat a^2 \hat \omega^1}{\hat r_+ \rho_H} \quad & n=1  \ , \\
\frac{\rho_H^{-n}}{\hat r_+}\Big[\hat r_+ \lambda^1_{\ell m} - 4 i \hat a^2 \hat \omega^1s 
 + \quad  &  \\
 -n 2 i\hat a^2 \hat \omega^1 \!\prescript{}{2}{F_1}\Big(1,1-n;2;\frac{\hat r_-}{\hat r_+}\Big)   \Big] \quad  & n>1 \ , \\
\end{cases}
\end{align}
The coefficients  $q^0_H(\hat r) ,p^0_H(\hat r), a^0_n$ and the boundary conditions for $\psi^{\textup{in},0}$ are given in Appendix~\ref{sec:HSC BCs} with $\omega \leftrightarrow\omega^0, \lambda_{\ell m \hat \omega}\leftrightarrow\lambda^0_{\ell m}$. 

For the solutions  $\psi^{\textup{up},0},\psi^{\textup{up},1}$, the system of differential equations is
\begin{align}
&\frac{d^2 \psi^{\textup{up},0}}{d\hat{r}^2} + p^0_\infty(\hat r)\frac{d\psi^{\textup{up},0}}{d\hat r} + q^0_\infty(\hat r)\psi^{\textup{up},0}=0 \, , \\
&\frac{d^2 \psi^{\textup{up},1}}{d\hat{r}^2} + p^0_\infty(\hat r) \frac{d\psi^{\textup{up},1}}{d\hat r} + p^1_\infty(\hat r)\frac{d\psi^{\textup{in},0}}{d\hat r} + \nonumber \\
&+ q^0_\infty(\hat r)\psi^{\textup{up},1} + q^1_\infty(\hat r)\psi^{\textup{up},0}=0 \, ,
\end{align}
where
\begin{align}
p^1_\infty(\hat r) &= -\frac{2\tilde G^1(\hat r;1)}{\Delta(\hat r^2 +\hat a^2)} \qquad  q^1_\infty(\hat r) = \frac{\tilde U^1(\hat r ;1)}{\Delta^2} \,, \\
\tilde G^1(\hat r;1) &=- i (\hat r^2 +\hat a^2)^2 \hat \omega^1 \,, \\
\tilde U^1(\hat r;1) &= -4 \hat \omega^1 [m \hat a (\hat r^2 + \hat a^2) +i (\hat r^2 -\hat a^2)s ] + \\
& - \Delta \Big(\lambda^1_{\ell m} + 2i\hat \omega^1\frac{\hat a^2}{\hat r} \Big) \, .
\end{align}
and the boundary conditions for $\psi^{\textup{up},1}$ are
\begin{equation}
\psi^{\textup{up},1}(\hat r)= \displaystyle \sum_{n=0}^{\infty} 
\frac{b^1_n}{\hat r^n} \ .
\end{equation}
The recursion relation for the coefficients $b^1_n$ is (setting $b^1_0=0$)
\begin{align}
b^1_n & = \frac{n-1}{2i\hat \omega^0}b^1_{n-1} + \displaystyle\sum_{k=1}^{n}\Big[q_\infty^{(k+1),0}- (n-k)p_\infty^{(k),0}\Big]\frac{b^1_k}{2i\hat \omega^0 n} + \nonumber \\
& +\displaystyle\sum_{k=1}^{n}\Big[q_\infty^{(k+1),1}- (n-k)p_\infty^{(k),1}\Big]\frac{b^1_k}{2i\hat \omega^0 n} - \frac{\hat \omega^1}{\hat \omega^0}b^0_n
\ ,
\end{align}
where 
\begin{align}
p_\infty^{(n),1} =
\begin{cases}
 2 i \hat \omega^1 \quad  & n=0 \ , \\[0.2em]
4 i \hat \omega^1 \quad & n=1  \ , \\[0.2em]
(4 i (\hat r_-^n - \hat r_+^n)\hat \omega^1\rho_H^{-1} \quad  & n>1 \ , 
\end{cases}
\end{align}
\begin{align}
q_\infty^{(n),1} =
\begin{cases}
0 \quad  & n=0,1 \ , \\
- \lambda^1_{\ell m}  -4 (\hat a\hat m + is)\hat \omega^1     \quad & n=1  \ , \\
\frac{2}{\rho_H}Q^1_1 + \frac{4 \hat \omega^1}{\rho_H^3}Q_2 \quad  & n>2 \ , 
\end{cases}
\end{align}
with
\begin{equation}
Q^1_1 = -\frac{1}{2}(\hat r_-^{n-1} -\hat r_+^{n-1} )\lambda^1_{\ell m}\ .
\end{equation}
The coefficients  $q^0_\infty(\hat r) ,p^0_\infty(\hat r), b^0_n$ and the boundary conditions for $\psi^{\textup{up},0}$ are given in Appendix~\ref{sec:HSC BCs} with $\omega \leftrightarrow\omega^0, \lambda_{\ell m \hat \omega}\leftrightarrow\lambda^0_{\ell m}$.  

\subsection{Linearization of the source} \label{sec:lin source terms}
In order to write the linearized amplitudes $Z^{H,\infty}_{\ell m \hat\omega} $ in the parameter $\sigma$, it is convenient first to recast Eq.~\eqref{eq:Z amp} as function of only $R^{\textup{in},\textup{up}}_{\ell m \hat \omega}$ and its first derivative. Taking advantage of the analyticity of the radial solutions in the positive real axis (except at the inner and outer horizons), second and higher order derivatives can be written solely in terms of $R^{\textup{in},\textup{up}}_{\ell m \hat \omega}$ and its first derivative. Thus, we can write Eq.~\eqref{eq:Z amp}  as
\begin{equation}
Z^{H,\infty}_{\ell m \hat{\omega}}  =\frac{2\pi }{W_{\hat r} }\bigg(X(\hat r) R^{\textup{in},\textup{up}}_{\ell m \hat \omega} +  Y(\hat r)\frac{\dd R^{\textup{in},\textup{up}}_{\ell m \hat \omega} }{\dd \hat r} \bigg)  \ , \label{eq:Z amp2}
\end{equation}
where $V(\hat r)$ is the Teukolsky potential of Eq.~\eqref{eq:radialTeueq}, while
\begin{align}
&X(\hat r) \equiv A_0 + \frac{V(\hat r)}{\Delta} C_2 - \frac{B_3}{\Delta}\frac{\dd V(\hat r)}{\dd \hat r}    \, , \\
&Y(\hat r) \equiv -C_1 + \frac{2(\hat r -1)}{\Delta}C_2  - \frac{B_3}{\Delta} (2 +V(\hat r))  \, ,\\
& C_1 \equiv A_1+B_1 \ ,\quad  C_2 \equiv A_2+B_2 \ \ .
\end{align}
After expanding Eq.~\eqref{eq:Z amp2} in the parameter $\sigma$, we can write the 0th order term as
\begin{equation}
Z^{\beta,0}_{\ell m}  =\frac{2\pi }{W^0_{\hat r} }\bigg(X^0(\hat r) R^{\alpha,0}_{\ell m} +  Y^0(\hat r)\frac{\dd R^{\alpha,0}_{\ell m}}{\dd \hat r} \bigg)  \ ,
\end{equation}
where $\beta = H (\infty)$ when $\alpha = \textup{in} (\textup{up})$, while
\begin{align}
&X^0(\hat r) \equiv A^0_0 + \frac{V(\hat r)}{\Delta}C^0_2  \, , \\[0.2em]
&Y^0(\hat r) \equiv -C^0_1 + \frac{2(\hat r -1)}{\Delta}C^0_2   \, ,\\[0.2em]
& V(\hat r) = -\frac{(K^0)^2 + 4i(\hat r-1)K^0}{\Delta} + 8i \hat \omega^0 \hat r + \lambda^0_{\ell m} \, ,\\[0.2em]
& K^0 = (\hat{r}^2 + \hat{a}^2)\hat \omega^0 -\hat a m \, , \\[0.2em]
&W^0_{\hat r} \equiv\frac{1}{\Delta}\left(\! R^{\textup{in},0}_{\ell m}\totder{R^{\textup{up},0}_{\ell m}}{\hat r} - R^{\textup{up},0}_{\ell m}\totder{R^{\textup{in},0}_{\ell m}}{\hat r} \!\right) \, .
\end{align}
Before writing the 0th order source terms $A^0_0, C^0_1, C^0_2$, we need to define the following  auxiliary quantities:
\begin{align}
S^0 &\equiv \prescript{}{-2}{S^0_{\ell m}}(\pi/2,c^0) \, ,\\
\tilde S^0 &= \totder{S^0}{\theta} - m S^0 + c^0 S^0   \, , \\
\mathcal{S}^0 &= -\frac{1}{2} S^0 \lambda^0_{\ell m} + \tilde S^0 \Big(c^0 - m -  \frac{i\hat a}{\hat r}\Big)   \, ,
\end{align}
and
\begin{align}
\mathcal{J}^0_z &= \tilde J^0_z -  \tilde E^0 \hat a \,, \\
P^0_\sigma &= - J^0_z \hat a +\tilde E^0 (\hat r^2 +\hat a^2) \, ,\\
\Gamma^0 &\equiv P^0_\sigma (\hat r^2 +\hat a^2) +\hat a \Delta \mathcal{J}^0_z \ .
\end{align}
The 0th order source terms can then be written as 
\begin{align}
A^0_0 &= - \frac{1}{2\hat r  \Gamma^0 \Delta} [\prescript{}{1}{A}^0_0 +  \prescript{}{2}{A}^0_0 + (\mathcal{J}^0_z)^2 S^0(\prescript{}{3}{A}^0_0 +\prescript{}{4}{A}^0_0)] \, ,\\
C^0_1 &=  \frac{\mathcal{J}^0_z}{\hat r\Gamma^0 } \big[i\hat r P^0_\sigma  \tilde S^0 +S^0 \mathcal{J}^0_z (\Delta + i \hat r^3 \omega^0 + i\hat a \hat r (c^0 -m)) \big] \, ,\\
C^0_2 &= \frac{S^0(\mathcal{J}^0_z)^2 \Delta}{2 \Gamma^0}\, ,
\end{align}
where
\begin{align}
\prescript{}{1}{A}^0_0 &=2 \hat r (P^0_\sigma)^2  \mathcal{S}^0 \, ,\\
\prescript{}{2}{A}^0_0 &=2P^0_\sigma \mathcal{S}^0 \mathcal{J}^0_z [(4 i  - m \hat a) \hat r +(\hat r^2 +\hat a^2)(\hat r \hat \omega^0 -2 i) ] \, ,\\
\prescript{}{3}{A}^0_0 &= 2 i (3 \hat a^2 \hat r + \hat r^3)\hat \omega^0 +(\hat r^2 + \hat a^2)^2 (\hat r \hat \omega^0 - 2i) \hat \omega^0 \, ,\\
\prescript{}{4}{A}^0_0 &= m\hat a^2 \hat r - 2 m\hat a[\hat a^2(\hat r \hat \omega^0 - i) +\hat r (3 i  - 2i \hat r + \hat \omega^0 \hat r^2 )] \,.
\end{align}
The 1th order correction $Z^{\beta,0}_{\ell m}$ is given by
\begin{align}
Z^{\beta,1}_{\ell m}  &=  \frac{2\pi }{W^0_{\hat r} }\bigg(X^1(\hat r) R^{\alpha,0}_{\ell m} +  Y^1(\hat r)\frac{\dd R^{\alpha,0}_{\ell m}}{\dd \hat r}  + \nonumber \\
&+ X^0(\hat r) R^{\alpha,1}_{\ell m} +  Y^0(\hat r)\frac{\dd R^{\alpha,1}_{\ell m}}{\dd \hat r}\bigg) -\frac{W^1_{\hat r}}{W^0_{\hat r}} Z^{\beta,0}_{\ell m} \ , 
\end{align}
where again $\beta = H (\infty)$ when $\alpha = \textup{in} (\textup{up})$, while
\begin{align}
&X^1(\hat r) \equiv A^1_0 + \frac{1}{\Delta}\Big(V^1(\hat r) C^0_2 + V^0(\hat r) C^1_2 - \totder{V^0(\hat r)}{\hat r}B^1_3\Big) \, , \\[0.2em]
&Y^1(\hat r) \equiv-C^1_1 + \frac{2(\hat r -1)}{\Delta}C^1_2  - \frac{2 + V^0(\hat r)}{\Delta}B^1_3  \, ,\\[0.2em]
& V^1(\hat r) = -\frac{2K^0 + 4i(\hat r-1)}{\Delta}K^1 + 8i \hat \omega^1 \hat r + \lambda^1_{\ell m} \, ,\\[0.2em]
& K^1 = (\hat r^2 +\hat a^2)\hat \omega^1 \, ,\\[0.2em]
&W^1_{\hat r} \equiv\frac{1}{\Delta}\left(\! R^{\textup{in},0}_{\ell m}\totder{R^{\textup{up},1}_{\ell m}}{\hat r} + R^{\textup{in},1}_{\ell m}\totder{R^{\textup{up},0}_{\ell m}}{\hat r} \!\right) + \nonumber \\
& \qquad - \frac{1}{\Delta}\left(\! R^{\textup{up},0}_{\ell m}\totder{R^{\textup{in},1}_{\ell m}}{\hat r} + R^{\textup{up},1}_{\ell m}\totder{R^{\textup{in},0}_{\ell m}}{\hat r} \!\right)\ .
\end{align}
The 1th order source terms  $A^1_0, C^1_1, C^1_2, A^1_3$ are quite cumbersome, and they are provided in a supplemental \textsc{Mathematica} notebook~\cite{webpage}.

Once the amplitudes $Z^{\beta,0}_{\ell m}, Z^{\beta,1}_{\ell m}$ with $\beta = (H, \infty)$ are known, it is possible to compute the corrections to the fluxes of Eqs.~\eqref{eq:flux corr inf} and \eqref{eq:flux corr hor} as follows
\begin{align}
I^0_{\ell m }(\hat r ,\hat \omega^0) &= \frac{\big|Z^{H,0}_{\ell m}\big|^2}{2\pi(\hat\omega^0)^2} \, ,\\
I^1_{\ell m }(\hat r ,\hat \omega^0,\hat \omega^1) &= \bigg(\frac{ Z^{H,0}_{\ell m }\bar Z^{H,1}_{\ell m}}{2\pi(\hat\omega^0)^2} + \text{c.c.} - 2\frac{\hat \omega^1}{\hat \omega^0}I^0_{\ell m }(\hat r ,\hat \omega^0)  \bigg) \, , \\
H^0_{\ell m }(\hat r ,\hat \omega^0) &= \frac{\tilde \alpha^0_{\ell m}}{2\pi} \big|Z^{\infty,0}_{\ell m}\big|^2 \, ,\\
H^1_{\ell m }(\hat r ,\hat \omega^0,\hat \omega^1) &=\frac{\tilde \alpha^0_{\ell m}}{2\pi}\Big( Z^{\infty,0}_{\ell m }\bar Z^{\infty,1}_{\ell m} + \text{c.c.}\Big)+\frac{\tilde \alpha^1_{\ell m}}{2\pi}\big|Z^{\infty,0}_{\ell m}\big|^2 \, ,
\end{align}
where $\text{c.c.}$ stands for complex conjugation, and
\begin{align}
\tilde \alpha^0_{\ell m} &= \frac{1}{\mathcal D^0}\big[256(2\hat{r}_+)^5\hat\kappa^0((\hat \kappa^0) ^2 + 4\epsilon^2)((\hat \kappa_0)^2 + 
16\epsilon^2)\hat \omega^0 \big] \, ,\\
\tilde \alpha^1_{\ell m} &= -\frac{\mathcal D^1}{\mathcal D^0} \tilde \alpha^0_{\ell m} + \frac{256(2\hat{r}_+)^5}{\mathcal C^0_{\ell m}}\hat \omega^1\big[ 64 \epsilon^4(\kappa^0 + \omega^0) +\nonumber \\
&+20 (\epsilon \kappa^0)^2 (\kappa^0 + 3\omega^0) + (\kappa^0)^4 (\kappa^0 +5 \omega^0) \big]\, , 
\end{align}
with $\epsilon = \sqrt{1-\hat{a}^2}/(4\hat{r}_+)$, $\hat \kappa^0 = \hat \omega^0 -\hat a m/(2\hat r_+) $ and
\begin{align}
\mathcal D^0 &= [(\lambda^0_{\ell m} + 2)^2 + 4 c^0(m- 
c^0)][(\lambda^0_{\ell m})^2 + 36 c^0(m- c^0)]\nn\\
 &+ (2\lambda^0_{\ell m}+3)[96(c^0)^2-48m c^0] +144(\hat \omega^0)^2(1-\hat a^2) \, , \\
\mathcal D^1 &= 4 \{( \lambda^0_{\ell m})^3 \lambda^1_{\ell m} + ( \lambda^0_{\ell m})^2[3 \lambda^1_{\ell m} + 10 (m -2 c^0) c^1] + \nonumber \\
&+ 2 \lambda^0_{\ell m} [\lambda^1_{\ell m} + 10 \lambda^1_{\ell m}c^0 (m -c^0) + 6 c^1(m + 2 c^0)] + \nonumber \\ 
&+72 \hat \omega^0\hat \omega^1 [1 +\hat a^2(m -2 c^0)(m -c^0)] + \nonumber \\
& +12 c^0 \lambda^1_{\ell m}(m+ c^0)\} \, .
\end{align}

\section{Assessment of the stability and convergence of the Fisher and covariance matrices} \label{app:Fisher}
In this appendix we provide some details on our procedure to assess the stability and numerical convergence of the Fisher and covariant matrices. 

This task is particularly delicate for EMRI waveforms, since the Fisher matrix is known to be ill-conditioned~\cite{Vallisneri:2007ev}.
In the best configuration, the condition number was $\kappa \sim 10^{12}$, while in worst scenario (typically occurring in the presence of a spinning secondary), the condition number was as large as $\kappa \sim 10^{20}$. Moreover, all waveform derivatives were computed numerically, which is an ill-conditioned operation.

To ameliorate the ill-condition issues, we performed our computation with arbitrary-precision arithmetic, obtaining Fisher matrices with precision no less than 38-digit in all elements and for all configurations.

We validated our Fisher analysis by:
\begin{itemize}
\item testing the stability of the Fisher and covariance matrices under random perturbations;
\item testing the convergence of the Fisher and covariance matrices under a change in the finite-difference parameter $\epsilon$ that regulates the accuracy of the numerical derivatives.
\end{itemize}

We check the stability of the Fisher and covariance matrices by perturbing each element with a deviation matrix $F^{ij}$. All elements of $F^{ij}$ are drawn from a uniform distribution $U$, which depends on the configuration under exam. Then, we compute
\begin{equation}
\delta_{\textit{\rm stability}} \equiv \underset{ij}{\max} \Bigg[ \frac{\big((\Gamma+F)^{-1}-\Gamma^{-1}\big)^{ij}}{(\Gamma^{-1})^{ij}}\Bigg]
\end{equation}

By performing a case-by-case careful analysis and boosting the numerical precision of our codes, we find that for the \emph{worst} cases in all configurations: 
\begin{itemize}
    \item the Fisher matrices converges within 2 orders of magnitudes in the $\epsilon$ parameters with relative deviations at the level of $0.03\%$ (another worst case is a convergence within 3 orders of magnitude in $\epsilon$ with deviations at $0.2\%$);
    \item the inverse matrix without priors converges in 2 order of magnitude in $\epsilon$  with deviations at $14\%$, while the diagonal elements converge with deviations at $0.1\%$;
    \item the inverse with priors converges in 2 order of magnitude in $\epsilon$ with deviations at $3.8\%$;
    \item the inverse without priors is stable with $\delta_{\textit{\rm stability}} = 7.5\%$ and perturbations $U[-10^{-7},10^{-7}]$;
    \item the inverse with priors is stable with  $\delta_{\textit{\rm stability}} = 4.1\%$ and perturbations $U[-10^{-6},10^{-6}]$.
\end{itemize}
Moreover, we noticed that, in order to achieve a convergent inverse with an accuracy of order $\mathcal O (1\%)$, it was necessary to compute a convergent Fisher matrices accurate at a the level of $\mathcal O (0.01\%)$.

Finally, it is worth noticing that, for some configurations in the presence of the secondary spin, we were unable to obtain a fully convergent covariance matrix: only the diagonal terms were convergent. Nonetheless, for all configurations presented in the main text the covariance matrix was found to be fully convergent.



\bibliographystyle{utphys}
\bibliography{Ref}

\end{document}